\documentclass[prc,twocolumn,epsfig,nofootinbib]{revtex4}
\usepackage{graphics}
\usepackage{epsfig}
\usepackage{amsfonts}
\usepackage{amsmath}
\usepackage{bm}

\usepackage{color}

\begin{document}
\title{Skyrme Random-Phase-Approximation description of lowest $K^{\pi}=2^+_{\gamma}$
states in axially deformed nuclei}
\author{
V.O. Nesterenko$^1$, V. G. Kartavenko$^1$, W. Kleinig $^{1,2}$,
J. Kvasil $^3$, A. Repko$^3$, R.V. Jolos$^1$, and P.-G. Reinhard$^4$}
\affiliation{$^1$ Laboratory of
Theoretical Physics, Joint Institute for Nuclear Research, Dubna, Moscow region, 141980, Russia}
\email{nester@theor.jinr.ru}
\affiliation{$^{2}$  \it Technische
Universit\"at Dresden, Institut f\"ur Analysis, D-01062, Dresden, Germany}
\affiliation{$^3$ Institute
of Particle and Nuclear Physics, Charles University, CZ-18000, Prague 8, Czech Republic}
\affiliation{$^4$ Institut f\"ur Theoretische Physik II,
Universit\"at Erlangen, D-91058, Erlangen, Germany}

\date{\today}

\begin{abstract}
The lowest quadrupole $\gamma$-vibrational $K^{\pi}=2^+$ states
in axially deformed rare-earth (Nd, Sm, Gd, Dy, Er, Yb, Hf, W) and actinide
(U) nuclei are systematically investigated within the
separable random-phase-approximation (SRPA) based on the Skyrme functional. The
energies $E_{\gamma}$ and reduced transition probabilities $B(E2)$ of
$2^+_{\gamma}$-states are calculated with the Skyrme forces SV-bas and SkM$^*$.
The energies of two-quasiparticle configurations forming
the SRPA basis are corrected by using the pairing blocking effect.
This results in a systematic  downshift of $E_{\gamma}$
by 0.3-0.5 MeV and thus in a better agreement with the
experiment, especially in Sm, Gd, Dy, Hf, and W regions. For other
isotopic chains, a noticeable overestimation of
$E_{\gamma}$ and too weak collectivity of $2^+_{\gamma}$-states still
persist.  It is shown that domains of nuclei with a low and high
$2^+_{\gamma}$-collectivity are related with the structure of the
lowest 2-quasiparticle states and conservation  of the  Nilsson selection
rules. The description of $2^+_{\gamma}$ states with SV-bas and
SkM$^*$ is similar in light rare-earth nuclei but deviates
in heavier nuclei. However SV-bas much
better reproduces the quadrupole deformation and energy of the
isoscalar giant quadrupole resonance. The accuracy of SRPA is
justified by comparison with exact RPA. The calculations suggest
that a further development of the self-consistent calculation schemes is needed
for a systematic satisfactory description of the $2^+_{\gamma}$ states.
\end{abstract}


\maketitle

\section{Introduction}
\label{sec:introduction}

During the last decades, remarkable progress was made in
description of nuclear dynamics within self-consistent mean-field (SCMF)
models (Skyrme, Gogny, relativistic), see e.g. the
reviews \cite{Ben03,Vre05,Pa07,Er10}.
In particular,  a variety of
quasiparticle random-phase-approximation (QRPA) methods
 was developed for the exploration
of small-amplitude excitations in deformed nuclei,
\cite{Ne06,Yosh08,Arte08,Losa10,Ter10,Terasaki_PRC_11,
Peru08,Peru11,Inakura09,Fracasso12,Hinohara13}.
So far these methods were mainly used for the description of
giant resonances (GR) in light
\cite{Yosh08,Arte08,Losa10,Ter10,Peru08,Inakura09,Fracasso12,Hinohara13}
and medium/heavy
\cite{Ne06,Ter10,Peru11,Fracasso12,IJMPE_08,kle_PRC_08,Kva_EPJA13,Ve09}
nuclei. However, self-consistent QRPA
was still rarely employed for the exploration  of the lowest vibrational states
($\beta$-, $\gamma$-, octupole) in deformed rare-earth and actinide
nuclei \cite{Terasaki_PRC_11,Peru11} (despite rich
available experimental information for these regions
\cite{bnl_exp,Sol_Grig}).
This is partly due to the huge configuration spaces required for such deformed heavy nuclei. However, the main problem lies in a high sensitivity of the
lowest vibrational states (LVS) to various factors. Following early calculations within the schematic Quasiparticle-Phonon Model (QPM)
\cite{Sol76,Sol_Shir,Sol_Sush_Shir}, the description of LVS requires a proper treatment of the single-particle (s-p) spectra near the Fermi level,
equilibrium deformation, pairing with the blocking effect, residual interaction (with both particle-hole and particle-particle channels), coupling to
complex configurations (with taking into account the Pauli principle), and exclusion of the spurious admixtures.  Besides, the description of LVS should be
consistent with the treatment of other collective modes, e.g. multipole GR.  All these factors and requirements make the self-consistent description of LVS
very demanding.

So far we are aware of two self-consistent QRPA studies of LVS in
rare-earth and actinide regions, one with Gogny forces for
$^{238}$U  \cite{Peru11} and another with Skyrme forces
for rare-earth nuclei \cite{Terasaki_PRC_11}.  Actually only the
last study \cite{Terasaki_PRC_11} is systematic. It covers
$\gamma$-vibrational $K^{\pi}=2^+_{\gamma}$ and $\beta$-vibrational
$K^{\pi}=0^+_{\beta}$ states in 27 rare-earth nuclei. The Skyrme
forces SkM$^*$ \cite{SkMs} and SLy4 \cite{SLy4} are used and
performance of SkM$^*$ is found noticeably better than of SLy4.
It is deduced that Skyrme QRPA is a reasonable basis for the
investigation of LVS.

In the present paper, we continue the systematic exploration of
$2^+_{\gamma}$-states in axial deformed nuclei with
QRPA using Skyrme forces. The $2^+_{\gamma}$-states
are chosen as the simplest case where we do not meet the problem of the
extraction of the spurious admixtures.  As compared to
\cite{Terasaki_PRC_11}, our study has some important new aspects.

First, it is desirable to use for description of $2^+_{\gamma}$-states
the Skyrme forces which simultaneously reproduce the energy of the
isoscalar giant quadrupole resonance (ISGQR). Following
\cite{IJMPE_08}, these forces should have a large isoscalar effective
mass $m^*_0/m$. The forces from \cite{Terasaki_PRC_11} have low
effective masses, $m^*_0/m=$0.70 for SLy4 \cite{SLy4} and 0.79 for
SkM$^*$ \cite{SkMs}, and so overestimate the ISGQR energy, see
\cite{IJMPE_08} and discussion below. To make the description of ISGQR
and $2^+_{\gamma}$-states consistent, we use in our calculations the
recent SV-bas force \cite{SV} with $m^*_0/m$=0.9. As shown below,
SV-bas also manages to reproduce systematically well ground state
deformations, a feature which is utterly crucial for a correct placing
of LVS. Note that very similar results were earlier obtained
\cite{nest_arXive15} with the Skyrme force SV-mas10 \cite{SV}
($m^*_0/m$=1.0).  We choose here SV-bas as a more general
parametrization which was already used in various studies, see
e.g. \cite{Er10,Kva_EPJA13,Ve09,Pot10}.  For comparison with
\cite{Terasaki_PRC_11}, the force SkM$^*$ is also implemented.

Second, we take into account the pairing blocking effect (PBE)
\cite{Sol76,Eisenberg-book,Ri80,Nilsson-book,BCS50} which, following
QPM studies \cite{Sol76,Sol_Shir,Sol_Sush_Shir}, can be important for
QRPA description of LVS in axially deformed nuclei.  The PBE weakens
the pairing and thus downshifts energies of low-energy
two-quasiparticle (2qp) states by a few hundreds keV
\cite{Sol76,Sol_Shir,Sol_Sush_Shir}, which in turn decreases the QRPA
energies of $2^+_{\gamma}$-states.  This effect can be especially
important for slightly collective states (with one or two dominant 2qp
components) which are often encountered amongst $2^+_{\gamma}$-states.
We implement PBE within the Bardeen-Cooper-Schrieffer (BCS) scheme
using volume pairing \cite{Ben00}. The same volume pairing, though in
the framework of the Hartree-Fock-Bogoliubov (HFB) approach without
PBE, was used in \cite{Terasaki_PRC_11}.

In fact, we are taking from the PBE only one aspect, namely the
modification of 2qp energies. The 2qp states as such (s.p. wave
functions and pairing occupation amplitudes) remain untouched.  This
ad hoc solution to the problem with the energies of
$2^+_{\gamma}$-states is admittedly not consistent. However it has a
great advantage not to disturb the orthonormality of 2qp basis and
thus allows to use the standard QRPA procedure.  At the same time,
following previous schematic \cite{Sol76,Sol_Shir,Sol_Sush_Shir} and
our present studies, the PBE for $2^+_{\gamma}$-states in medium and
heavy deformed nuclei can be strong and certainly deserves the
consideration.  In this connection, our PBE-QRPA calculations can be
viewed as a first step highlighting the problem and calling for
further checking within a self-consistent PBE-QRPA prescription, yet
to be developed.

The third new aspect is that we provide a detailed analysis of the
obtained results, both numerically and analytically (e.g. in terms of
simplified models for schematic RPA).  We determine domains of nuclei
with low and high collectivity of $2^+_{\gamma}$-states and
demonstrate that the lowest $K^{\pi}=2^+$ 2qp state plays a key role
in formation of these domains.  The study embraces 9 isotopic chains
(Nd, Sm, Gd, Dy, Er, Yb, Hf, W, U) with 41 axially deformed nuclei, as
compared to 27 rare-earth nuclei in \cite{Terasaki_PRC_11}.

The calculations are performed within the separable
random-phase-approximation (SRPA) method based on the Skyrme
functional \cite{Ben03,Skyrme,Vau}. The method is  developed in a
one-dimensional (1D) version for spherical nuclei \cite{Ne02} and a
two-dimensional (2D) version \cite{Ne06,Ne_ar05} for axial deformed nuclei.
SRPA is derived self-consistently:
i) both the mean field and
residual interaction are obtained from the same Skyrme functional, ii)
the residual interaction includes all terms of the Skyrme functional as well as
the Coulomb (direct and exchange) terms.  The self-consistent
factorization of the residual interaction dramatically reduces the
computational effort for deformed nuclei while keeping high accuracy
of the method. However SRPA is not self-consistent in the part of the pairing
interaction because i) ad hoc implementation of PBE into QRPA and ii)
skipping the particle-particle channel in the residual interaction.

In earlier studies, SRPA was successfully applied for the description
of various GR in spherical and deformed nuclei: E1(T=1) and E2(T=0)
\cite{Ne06,IJMPE_08,Ne02,kle_PRC_08}, toroidal/compression E1
\cite{Kva_EPJA13}, and spin-flip M1 \cite{Ve09}).  However, the success of
the model for GR does not mean that it is also robust in description
of so fragile excitations as LVS.  In this connection, we compare
below some SRPA results with those obtained with the exact (no the
separable ansatz) 2D QRPA code \cite{repko}. We find a nice agreement
which confirms that SRPA is accurate enough to pretend for description
of $2^+_{\gamma}$ states.

The paper is organized as follows. In Sec. 2, method and calculational
details are outlined. The equations for the pairing blocking are
given, the SRPA scheme is sketched, and SRPA results are compared with
those from the exact QRPA.  It is shown that SV-bas, unlike SkM*,
nicely reproduces equilibrium quadrupole deformations and the ISQGR
energy.  Sec. 3 presents the main results for energies and reduced
transition probabilities B(E2) of $2^+_{\gamma}$-states. In Sec. 4,
these results are discussed and analyzed in detail and compared with
the previous data \cite{Terasaki_PRC_11}. A summary is given in
Sec. 5. In Appendix A, the expression for the pairing matrix element
is derived.  In Appendix B, the basic SRPA equations are outlined. In
Appendix C, a simple two-pole RPA model is presented to be applied for
explanation of the domains with low and high collectivity of
$2^+_{\gamma}$-states. In Appendix D, SRPA strength constants of the
residual interaction are compared with those of the QPM.

\section{Model and calculation scheme}

The SRPA approach \cite{Ne06} used in this paper is based on
the Skyrme functional \cite{Ben03}
\begin{equation}
\mathcal{E}(\rho,\tau, \bf{J}, \bf{j}, \bf{\sigma}, \bf{T}) =
\mathcal{E}_{\rm{kin}} + \mathcal{E}_{\rm{Sk}} +
\mathcal{E}_{\rm{Coul}} + \mathcal{E}_{\rm{pair}}
\label{1}
\end{equation}
where $\mathcal{E}_{\rm{kin}}$ is the kinetic energy,
$\mathcal{E}_{\rm{Sk}}$ is the potential energy according to the
Skyrme functional, $\mathcal{E}_{\rm{Coul}}$ is the Coulomb energy,
and $\mathcal{E}_{\rm{pair}}$ is the pairing energy.  The Coulomb
exchange term is treated in Slater approximation. The volume pairing
corresponds to a zero-range pairing interaction.  The Skyrme part
$\mathcal{E}_{\rm{Sk}}$ depends on the local densities and currents:
density $\rho(\mathbf{r})$, kinetic-energy density $\tau(\mathbf{r})$,
spin-orbit density $\mathbf{J}(\mathbf{r})$, current
$\mathbf{j}(\mathbf{r})$, spin-density $\mathbf{\sigma}(\mathbf{r})$,
and spin-kinetic-energy density $\mathbf{T}(\mathbf{r})$
\cite{Ben03}.  The mean-field Hamiltonian and SRPA residual
interaction are self-consistently determined through the first and
second functional derivatives of (\ref{1}), respectively \cite{Ne06}.
Further details of the model and calculation scheme are given below.

\subsection{Mean field and quadrupole deformation}

The stationary 2D mean-field calculations are performed with the SKYAX
code \cite{SKYAX} in cylindrical coordinates using a mesh size of 0.5
fm and a box size of
about three nuclear radii. The single-particle space is chosen to
embrace the levels from the bottom of the potential well up to energy
15-20 MeV. For SV-bas, the s-p schemes involve  304
proton and 375 neutron levels in $^{150}$Nd and  379 proton
and 485 neutron levels in $^{238}$U.

The ground state is obtained by solving the mean-field equations
and resides at the minimum of the total energy (\ref{1}).  Its axial
quadrupole deformation is characterized by the dimensionless
deformation parameter \cite{Raman}
\begin{equation}\label{eq:quad_def}
  \beta_2
  =
  \sqrt{\frac{5\pi}{3}} \frac{Q_2}{Z R^2} ,
\end{equation}
where $Q_{2} = \int d{\mathbf{r}} \rho_p({\mathbf{r}}) r^2 Y_{20}$
is the quadrupole moment, $R = 1.2\;\rm{fm} A^{1/3}$, A is the mass number.

\begin{figure}[t]
\includegraphics[width=8.5cm]{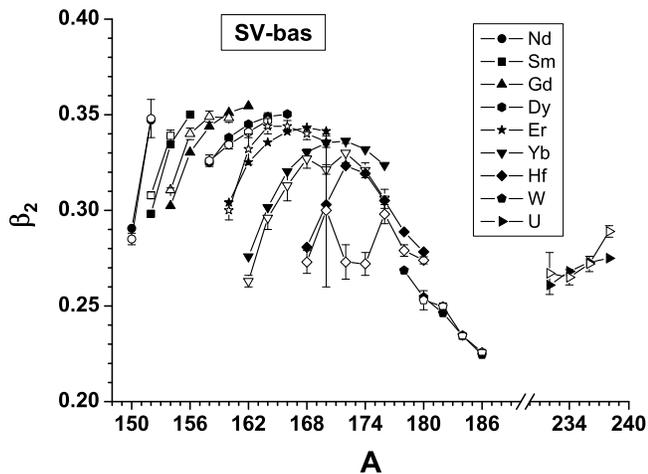}
\caption{\label{fig:deform1}Parameter $\beta_2$ of the axial
  quadrupole deformation in rare-earth and actinide nuclei. The values
  calculated with SV-bas \protect\cite{SV} (full
  symbols) are compared with the experimental data
  \protect\cite{bnl_exp}
   (open symbols with error bars).}
\end{figure}

\begin{figure}[t]
\includegraphics[width=8.5cm]{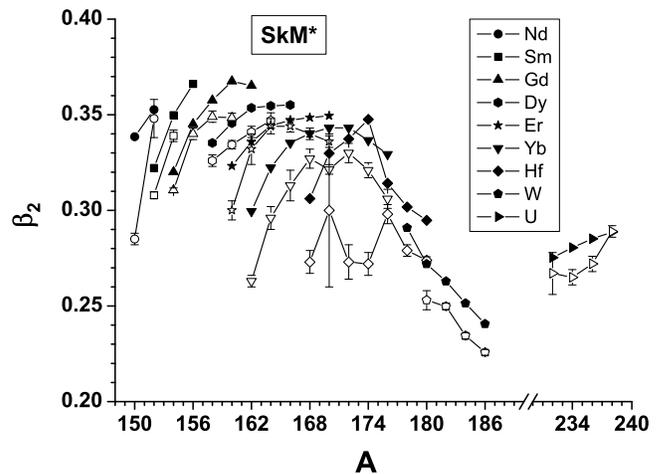}
\caption{\label{fig:deform2}The same as Fig. \ref{fig:deform1} but for SkM$^*$.}
\end{figure}

Figure \ref{fig:deform1} compares deformation parameters calculated
using SV-bas with available experimental data \cite{bnl_exp} and
Fig. \ref{fig:deform2} shows the same comparison for SkM$^*$.  Figures
1 and 2 show very nice agreement for SV-bas while SkM$^*$
systematically overestimates $\beta_2$, especially in Yb, Hf, W, and U
isotopes.  Note that both SV-bas and SkM$^*$ fail to describe the
specifically low values of experimental $\beta_2$ in $^{170}$Yb and
$^{172,174}$Hf. Note also the exceptionally large error bars in
$^{170}$Hf.


\subsection{Pairing and blocking effect}

The volume pairing interaction reads
\begin{equation}
V^q_{\rm{pair}} ({\mathbf{r}}, {\mathbf{r}}') =
V_{q}
\: \delta(\mathbf{r} - \mathbf{r}') ,
\label{5}
\end{equation}
where $q$ stands for protons or neutrons and $V_{q}$ are pairing
strengths.  In the present study pairing is treated at the
BCS level \cite{Ben00}.

If the pairing-blocking effect (PBE) is accounted for, the BCS problem
is solved separately for the ground $\Psi_0^q$ and excited
n-quasiparticle $\Psi_n^q$ states. For the ground state, the
expectation value
$\langle \Psi_0^q |H_{\rm{pair}}|\Psi_0^q \rangle$
for the pairing Hamiltonian $H_{\rm{pair}}$ is minimized to determine
the set of Bogoliubov coefficients $\{u_k^q, v_k^q\}$.  For
$n$-quasiparticle excitation, the wave function reads
\begin{eqnarray}
\label{nqp}
\Psi_n^q &=&\hat\alpha^+_{j_1} ... \hat\alpha^+_{j_n} \Psi_0^q =
\hat{a}^+_{j_1} ... \hat{a}^+_{j_n} \\ &\cdot& \prod_{k \ne j_1 ...j_n
  \in q} [u_k^q(j_1 ...j_n)+ v_k^q(j_1 ...j_n) \hat{a}^+_{k}
  \hat{a}^+_{\bar{k}}] |-\rangle, \nonumber
\end{eqnarray}
where $ \hat{a}^+_{j} (\hat{\alpha}^+_{j})$ creates the particle
(quasiparticle) at the state $j$ and $|-\rangle$ is the particle
vacuum.  For this excitation, the expectation $\langle \Psi_n^q
|H_{\rm{pair}}|\Psi_n^q \rangle$ is minimized and the new set of
occupation numbers $\{u_k^q(j_1,..j_n), v_k^q(j_1,..j_n)\}$ specific
for the given excitation is determined. In the latter case, the BCS
equations for axially deformed nuclei (with doubly degenerate s-p
levels) have a peculiarity: if some states from the set $\{j_1,
... j_n\}$ are unpaired, then these states are excluded from the
pairing scheme and contribute to the BCS equations as pure
single-particle states. The physics behind is obvious: if some level
is occupied by an unpaired nucleon, then it is closed (=blocked) for
the pairing process which transfers nuclear pairs.  This is why it is
called pairing blocking effect
\cite{Sol76,Eisenberg-book,Ri80,Nilsson-book,BCS50}.

The PBE takes place in both BCS and HFB theories as soon as we deal
with n-quasiparticle excitations. Most often the PBE is considered for
1qp excitations in odd and odd-odd nuclei, see
e.g. \cite{Ben00,Dug01,Pot10} and more references in
\cite{BCS50}. Following QPM studies
\cite{Sol76,Sol_Shir,Sol_Sush_Shir}, the PBE may play a role in QRPA
description of LVS in even-even axially deformed nuclei.  Indeed 2qp
states constitute the configuration space for QRPA. The first
low-energy 2qp states are the main contributors to the lowest QRPA
excitation. So it is worth to check how PBE for the low-energy 2qp
states affects the description of LVS.

The main effect of the PBE is to change the 2qp energies
\cite{Sol76,Sol_Shir,Sol_Sush_Shir}.  Thus we use here in ad-hoc
manner only one PBE output, PBE-corrected 2qp energies.  Only them
are implemented to QRPA while the occupation amplitudes ($u,v$) and
s.p. wave functions are kept the same as in the BCS ground state. This
has the advantage that orthonormality of the 2qp configuration space
is maintained and the standard QRPA scheme remains applicable.

Usually in BCS+QRPA calculations the 2qp-energies are computed
by using the pairing gaps
$\Delta_q$, chemical potentials $\lambda_q$ and Bogoliubov
coefficients $\{u_k^q, v_k^q\} \in q$ for the ground BCS state,
yielding
\begin{equation}
\label{eps_ij}
\epsilon_{ij}^q=\epsilon_i^q +\epsilon_j^q
\end{equation}
where $\epsilon_i^q = \sqrt{(\tilde{e}_i^q - \lambda_q)^2 +
  \Delta_q^2}$ is the energy of the 1qp state, $\tilde{e}_i^q $ is the
renormalized s-p energy (see the expression below). In HFB+QRPA
calculations, the 2qp states for the QRPA configuration space are also
expressed in terms of ground state values.
In particular, their energies are calculated as a sum
of two 1qp energies in the canonical basis using the HFB solutions for
the ground state, see e.g. \cite{Ter10,Terasaki_PRC_11,Peru11}.  Both
such BCS and HFB schemes do not include the PBE for the 2qp
states. Following QPM and our calculations, such a treatment can be
insufficient for a correct description of the LVS.

For $K^{\pi}=2^+$ states
\begin{eqnarray}
\label{n2p}
\Psi^q(ij)&=&\hat\alpha^+_{i}\hat\alpha^+_{j} \Psi_0^q
\\
&=& \hat{a}^+_{i}\hat{a}^+_{j}
\prod_{k \ne i,j \in q}
(u_k^q(ij)+ v_k^q(ij) \hat{a}^+_{k} \hat{a}^+_{\bar{k}}) |-\rangle,
\nonumber
\end{eqnarray}
the 2qp pairs are necessarily non-diagonal ($i \ne j$).
For a {\it constant} pairing force, the BCS-PBE prescription for this
case was formulated in \cite{Sol76}. Below we present the BCS-PBE
formalism for the $\delta$-force volume pairing (\ref{5}). For each
2qp state $\Psi^q(ij)$, one should solve the system of BCS+PBE
equations
\begin{equation}
[u_k^q(ij)]^2=
  \frac{1}{2} \: \left\{1 + \frac{\tilde{e}_k^q -
  \lambda_q(ij)}{\sqrt{\left[\tilde{e}_k^q - \lambda_q(ij)\right]^2
  + [\Delta_k^{q}(ij)]^2}} \right\},
\label{32}
\end{equation}
\begin{equation}
 [v^q_k(ij)]^2 =
  \frac{1}{2} \: \left\{1 - \frac{\tilde{e}_k^q -
  \lambda_{q}(ij)}{\sqrt{\left[\tilde{e}_k^q - \lambda_{q}(ij)\right]^2
  + [\Delta_k^{q}(ij)]^2}} \right\},
\label{33}
\end{equation}
\begin{equation}
  \Delta_k^{q}(ij)
  = -  \sum_{k'\neq i,j}^{K'>0, k' \in q}
  f_{k'}^q V^{(\rm{pair, q})}_{k \bar{k} k' \bar{k'}} \: v_{k'}^q (ij) u_{k'}^q (ij),
\label{34}
\end{equation}
\begin{equation}
  N_{q}
  =
  2 + \sum_{k'\neq i,j}^{K'>0, k' \in q} f_{k'}^q
  \left\{1 - \frac{\tilde{e}_k^q - \lambda_{q}(ij)}{\sqrt{\left[\tilde{e}_k^q -
         \lambda_{q}(ij)\right]^2 + [\Delta_k^{q}(ij)]^2}} \right\} ,
\label{35}
\end{equation}
where
\begin{equation}
\tilde{e}_k^q=e_k^q-1/2\sum_{k'\in q} f_{k'}^q V^{(\rm{pair, q})}_{k \bar{k} k' \bar{k'}} [v_k^q]^2
\end{equation}
is the renormalized s-p energy and $e_k^q$ is the initial s-p energy.
Furthermore,
$u_{k}^q(ij), v_{k}^q(ij), \Delta_k^{q}(ij), \lambda_{q}(ij)$ are
Bogoliubov coefficients, pairing gaps and chemical potentials,
calculated for the 2qp $(ij)$-excitation. The sums in (\ref{34}) and
(\ref{35}) include all s-p states $k'$ (with
isospin $q$ and projection $K'>0$ of the total angular
momentum)
for exception of $k'=i$ and $j$;
$N_p=Z$ and $N_n=N$ are proton and neutron numbers.  The
smoothing energy-dependent cut-off weights $f_{k'}^q$ are introduced to
cure the well-known drawback of the zero-range pairing force to
overestimate the coupling to the (continuum) high-energy states
\cite{Ri80,BCS50}. Expressions for weights $f_{k'}^q$ and pairing matrix
elements $V_{k \bar{k} k' \bar{k'}}^{(\rm{pair, q})}$ in axial
nuclei are given in the Appendix A.

The PBE-corrected energy of the 2qp excitation reads
\begin{equation}
  \mathcal{E}^{q}_{\rm{bl}}(ij)
  =
  \mathcal{E}^{q}(ij) - \mathcal{E}^{q}_0
\label{37}
\end{equation}
where
\begin{eqnarray}
\label{38}
  &&
   \mathcal{E}^{q}_{\rm{bl}}(ij)=\langle \Psi^q(ij) |H^q_{\rm{pair}}|\Psi^q(ij)\rangle
  =
   \tilde{e}^q_i + \tilde{e}^q_j +
\\
  && \quad
  + \sum_{k \neq i,j}^{K>0, k \in q}
  f^q_k[2 \tilde{e}^q_k \: (v^q_k(ij))^2 -\Delta_k^{(q)}(ij) \: u^q_k(ij) \: v^q_k(ij)]
\nonumber
\end{eqnarray}
is the energy of the system in the (ij)-state and
\begin{eqnarray}
\label{30}
  \mathcal{E}^{q}_0&=&\langle \Psi^q_0 |H^q_{\rm{pair}}|\Psi^q_0\rangle
\\
  &=&2\sum_{k}^{K>0, k \in q} f^q_k \tilde{e}^q_k \: (v^q_k)^2
  -\sum_{k}^{K>0, k \in q} f^q_k
  \Delta^q_k \: u^q_k \: v^q_k \;,
\nonumber
\end{eqnarray}
is the energy of the q-subsystem in the ground state.
The values $u_{k}, v_{k}, \Delta_k^{q}, \lambda_{q}$
in (\ref{30}) are for the ground state.
Eqs. (\ref{34}), (\ref{35}), (\ref{38}) show that PBE excludes
the states $i$ and $j$ from the pairing sums.  These blocked states do
not contribute to the pairing gap (\ref{34}) and enter (\ref{35}) and
(\ref{38}) as single-particle (not quasi-particle) states.

The sums in (\ref{34}), (\ref{35}), and (\ref{38}) are usually
dominated by a few $k'$-states around the Fermi level. If the states
$i$ and $j$ belong to this group, then their blocking can effectively
decrease the level density near the Fermi level and thus the pairing
gap (\ref{34}). Consequently the energy (\ref{38}) is changed.  In
such cases, the pairing is significantly suppressed and the BCS-PBE
value for the 2qp energy(\ref{37}) becomes a few hundreds of keV
smaller than the BCS energy (\ref{eps_ij}) \cite{Sol76}. This in turn
leads to a significant downshift of the energy of the first QRPA
solution.

In the present study, we block the five lowest $K^{\pi}=2^+$ 2qp
states (proton and neutron altogether). The calculations show that
this number of blocked states is optimal. More blocking would involve
the states remote by energy from the Fermi level and thus with a
negligible PBE.  Less blocking is likely to miss a part of the PBE
corrections.

We substitute the PBE-corrected energies
$\mathcal{E}^{q}_{\rm{bl}}(ij)$ to SRPA replacing the
$\epsilon^q_{ij}$. However we do not use the PBE modified Bogoliubov
coefficients $\{u^q_{k}(ij), v^q_{k}(ij)\}$. Instead we continue to
employ in QRPA the ground state set $\{u^q_{k}, v^q_{k}\}$ and wave
functions. This leaves the 2qp basis orthonormalized and renders our
PBE-SRPA scheme easily applicable.

It is also worth to inspect a possible impact of our scheme
on the basic features of QRPA, namely
stability of the QRPA interaction matrix, elimination of spurious
modes, and sum rules: i) Concerning the QRPA matrix,
the PBE-induced reduction of the positive diagonal elements (2qp energies)
of the matrix indeed can cause instabilities in some cases.
This is checked numerically. We find that for the
$K^{\pi}=2^+$ states studied here the QRPA remains in the stable regime.
The only exception is $^{164}$Dy in
the calculations with the force SkM*, see discussion below.
ii) Spurious modes must be carefully checked  when trying
to apply the PBE to other quadrupole states, say with $K^{\pi}=0^+$
and $K^{\pi}=1^+$, but not in our case. For $K^{\pi}=2^+$ states  considered
in the present study the spurious modes are absent at all.
iii) Concerning the sum rules, there is some quantitative effect. But it is
 extremely small as the main contribution to sum rules comes from
higher lying states which are not affected by the PBE.
Altogether, the present ad hoc implementation of the PBE looks
robust. It still calls for a thorough formal self-consistent
development which, however, will be tedious and take time. We consider
the present study as a first step in exploration of the impact of the
PBE on low-lying spectra of $K^{\pi}=2^+$ states.

The PBE should be applied with care in case of a weak pairing because
the blocking reduces pairing and may trigger its full break-down. In
the worst case, a more involved formalism (allowing a weak pairing)
should be used, e.g. the method with particle-number projection before
variation \cite{Kuz}. Calculations with this method show that BCS-PBE
somewhat underestimates the 2qp energies \cite{Kuz}.  However, the
projection method requires a huge effort and it cannot be consistently
applied for Skyrme energy functional \cite{Dob07}.  So we use here
BCS-PBE, though with staying alert for suspect cases.

\subsection{SRPA scheme}

The SRPA formalism for axial nuclei is described in detail elsewhere
\cite{Ne06,Ne_ar05}.  Here we sketch only the points relevant for the
present study.  As mentioned above, the SRPA formalism starts from the
functional (\ref{1}).
The
residual interaction includes contributions from both time-even and
time-odd densities and also takes care of the Coulomb interaction.
The coupling between the quadrupole $\lambda\mu$=22 and hexadecapole
$\lambda\mu$=42 modes, pertinent to deformed nuclei, is included.
The basic SRPA equations and more calculation details can be found in the
Appendix B.

The present SRPA version skips the particle-particle (hole-hole)
channel for $K^{\pi}=2^+$ states. In
QPM the pp-channel is used to harmonize description of LVS energies
and transition probabilities \cite{Sol_Sush_Shir} but these
calculations are not self-consistent. The self-consistent Skyrme
BCS-QRPA calculations for spherical nuclei show that the pp-channel
tends to decrease the LVS energies \cite{Sev08}. If so, then this
effect can be partly compensated by the energy upshift gained by using
the particle-projection method \cite{Kuz}. The Skyrme HFB-QRPA studies
of LVS in deformed nuclei use the pp-channel only partly
\cite{Terasaki_PRC_11} if at all \cite{Ter10}. In general, the
pp-channel, being crucial for $\beta$-vibrational $K^{\pi}=0^+$
states, seems not to be so important for $\gamma$-vibrational
$K^{\pi}=2^+$ states. At least we do not know any self-consistent
study for the lowest $K^{\pi}=2^+$ states in axial deformed nuclei,
which would demonstrate a real need for this channel.

In the present study, we calculate the structure and energies of the
first RPA one-phonon $2^+_{\gamma}$ states ($\lambda\mu\nu=221$) in
Nd, Sm, Gd, Dy, Er, Yb, Hf, W, and U isotopes. The reduced probability
B(E2)=$|\langle\nu=1|\sum^Z_{k=1} r_k^2 Y_{22}(\theta_k)|0\rangle|^2$
of the transition from the ground $|0\rangle$ to the SRPA $\nu=1$
state are also computed.
\begin{figure}[t]
\includegraphics[width=8cm]{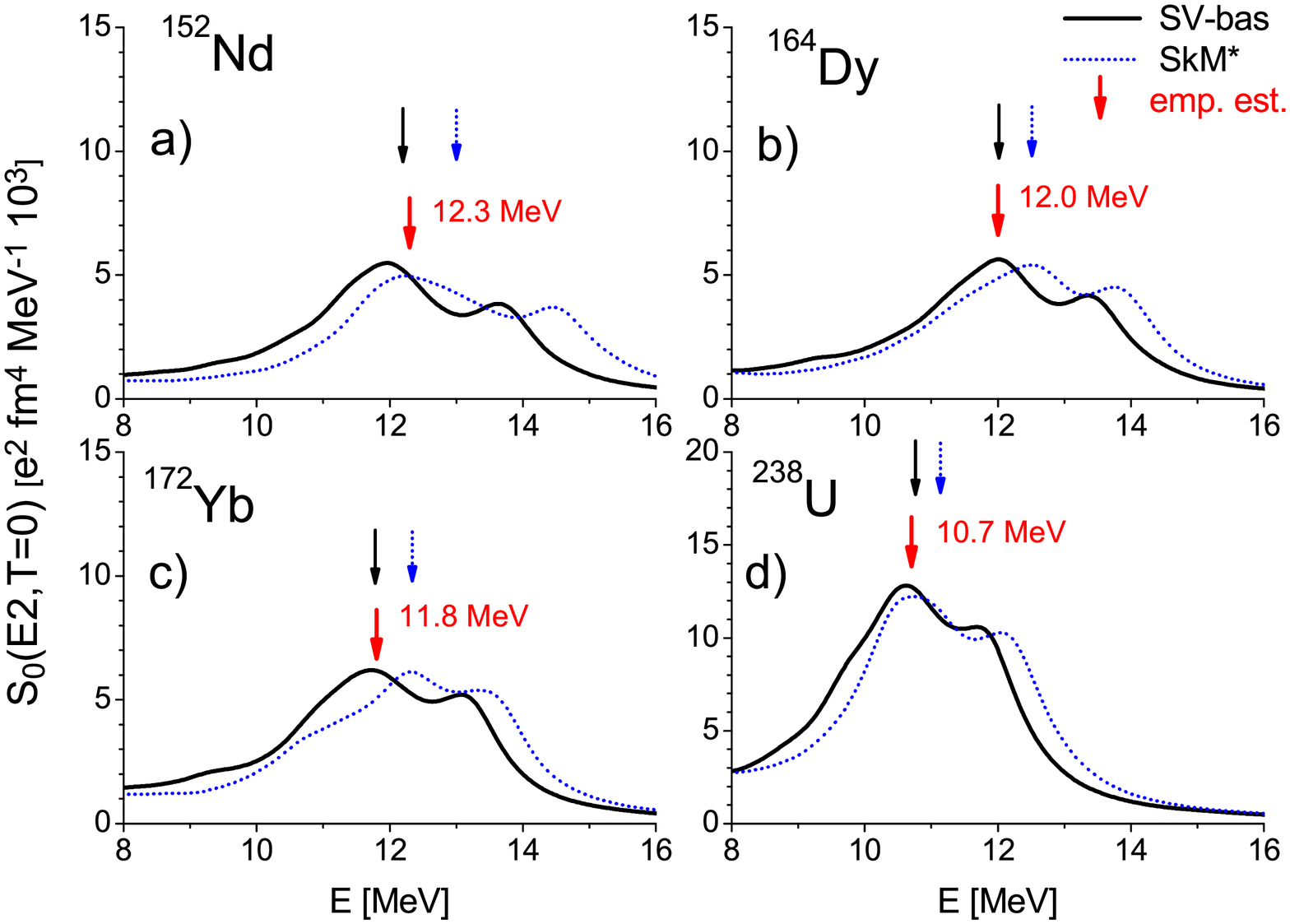}
\caption{(Color online) The isoscalar strength function for the ISGQR in $^{152}$Nd
$^{164}$Dy, $^{172}$Yb and $^{238}$U, calculated with the Skyrme forces SkM$^*$
\protect\cite{SkMs} (dotted blue line) and SV-mas10  \protect\cite{SV} (solid black line).
The Lorentz averaging parameter is $\Delta$ = 1 MeV. The empirical estimations
for the ISGQR centroids \cite{Scamps_PRC_13} are marked by lower red arrows with indicated energies.
The SV-bas and SkM$^*$ estimations for the centroids are denoted by upper black solid and blue dotted
arrows, respectively.}
\end{figure}

The configuration space for $\lambda\mu=22$ involves, depending on the
isotope, 6600-9600 proton and 9400-14200 neutron 2qp-states with
excitation energies up to 55-80 MeV. This basis is sufficient for our
aims. It results (together with the quadrupole components $\lambda\mu$=20
and 21) in a reasonable exhaustion of the total energy-weighted sum
rule EWSR(E2,T=0)= $(\hbar^2e^2)/(8\pi m_p) 50 A \langle r^2
\rangle_A$ by $\sim 95-98\%$. A similar size of configuration space
was used in \cite{Terasaki_PRC_11} and \cite{Peru11} (19000-28000 and
23000-26000 2qp states, respectively).

The calculations are performed for the Skyrme parametrizations
SV-bas and SkM$^*$.  As mentioned in the introduction, SV-bas is
chosen because it provides an accurate description of the
ground state deformations and ISGQR energies.  This is demonstrated
in Fig. 3, where ISGQR strength functions and energy
centroids (see definitions in the Appendix B) are depicted
for SV-bas and SkM$^*$.
The calculated centroids
are 12.2 and 13.0 MeV in $^{152}$Nd, 12.0 and 12.5 MeV in $^{164}$Dy, 11.8 and 12.3 MeV
in $^{172}$Yb, and 10.7 and 11.1 MeV in $^{152}$U, for SV-bas and SkM$^*$ respectively.
These results are compared with the empirical polynomial estimations
\cite{Scamps_PRC_13}.  It is seen that SV-bas well describes the  energy
centroids while SkM$^*$ systematically overestimates them.
So SV-bas demonstrates a good reproduction of both
axial deformations and ISGQR
energies which makes SV-bas a promising candidate for the description of
$\gamma$-vibrational states.

\begin{figure}[t]
\includegraphics[width=8cm]{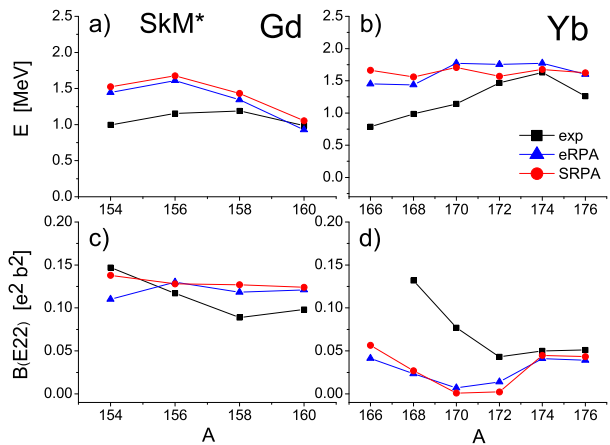}
\caption{(Color online)
Energies (a-b) and B(E2) values (c-d) of the $2^+_{\gamma}$-vibrational states,
calculated with the force SkM$^*$ in the framework of SRPA (red circles) and exact eRPA
(blue triangles) in Gd (left) and Yb (right) isotopes. In both calculations,
the PBE and pp-channel in the residual interaction
are omitted. The experimental data \protect\cite{bnl_exp} are depicted by black squares.}
\end{figure}

To demonstrate the accuracy of SRPA, we compare in Fig. 4 some results
for $K^{\pi}=2^+_{\gamma}$ states obtained within SRPA and exact 2D
QRPA \cite{repko}.  The exact method is noted as eRPA. In both cases,
the calculations are performed without PBE and pp-channel in the
residual interaction. The isotopic chains with a high (Gd) and low
(Yb) collectivity of $2^+_{\gamma}$ states are considered.  We see a
very nice agreement between SRPA and eRPA results, which
demonstrates the robustness of SRPA. Since SRPA calculations
require much less computational effort than eRPA, just SRPA is used in
the following.
\begin{figure*}[t]
\includegraphics[width=11.5cm]{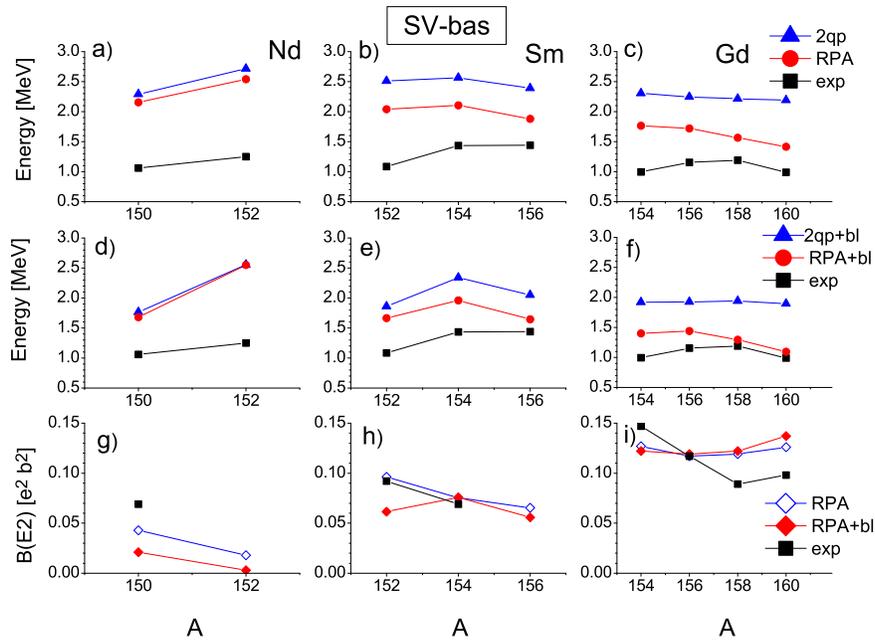}
\caption{(Color online) The lowest 2qp and QRPA(RPA) energies (a-f) as
  well as B(E2) values (g-i) of 2$^+_{\gamma}$-vibrational states in
  Nd (left), Sm (center) and Gd (right) isotopes, calculated with the
  force SV-bas. The 2qp (filled blue
  triangles) and QRPA(RPA) (filled red circles) energies are obtained without
  (a-c) and with (d-f) PBE. The QRPA(RPA) B(E2) values without (empty blue
  diamonds) and with (filled red diamonds) PBE are plotted in (g-i).
  In all the plots, the experimental data \protect\cite{bnl_exp} are given
  (filled black squares).}
\end{figure*}
\begin{figure*}
\includegraphics[width=12.0cm]{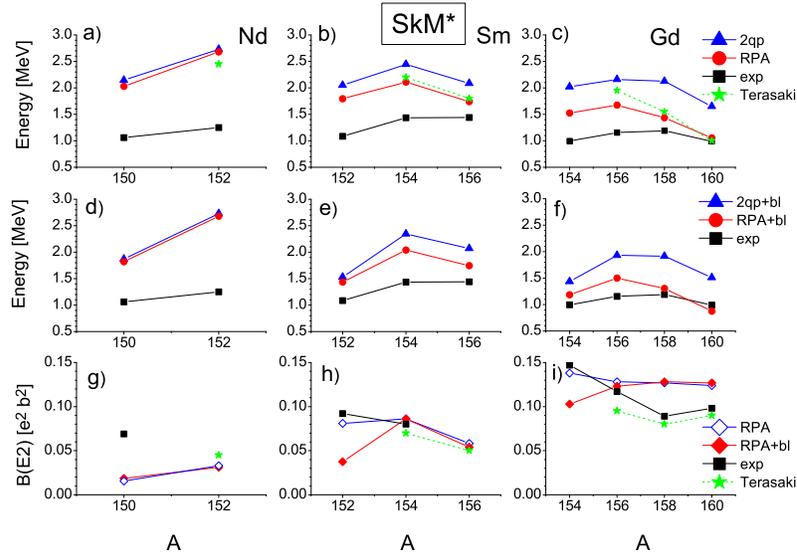}
\caption{(Color online) The same as in Fig. 5 but for SkM$^*$.
For the comparison, the SkM$^*$ results \protect\cite{Terasaki_PRC_11} are
depicted (filled green stars).}
\end{figure*}

\section{Main results}
\subsection{Main results}

Results of our calculations for the lowest 2qp states, QRPA energies,
and B(E2)-values of $2^+_{\gamma}$ states are presented in
Figs. 5-10. Cases without and with PBE are considered using for 2qp
energies Eqs.  (\ref{eps_ij}) and (\ref{37}), respectively. The
results are compared with available experimental data \cite{bnl_exp}.
Note that experimental errors for $2^+_{\gamma}$ energies are
typically $\pm$ 0.01 MeV, i.e. much smaller than the relevant values to be
discussed. Concerning B(E2), the errors usually do not exceed 10$\%$ for
collective states (B(E2)$>$0.1-0.09 $e^2b^2$) but can reach 15-30$\%$ in
less collective states ($^{150}$Nd, $^{154}$Sm, $^{170-176}$Yb, $^{238}$U).
In Figs. for SkM*, results are compared with those of \cite{Terasaki_PRC_11}
(manually extracted from the figures of the paper).

\begin{figure*}[t]
\includegraphics[width=12.5cm]{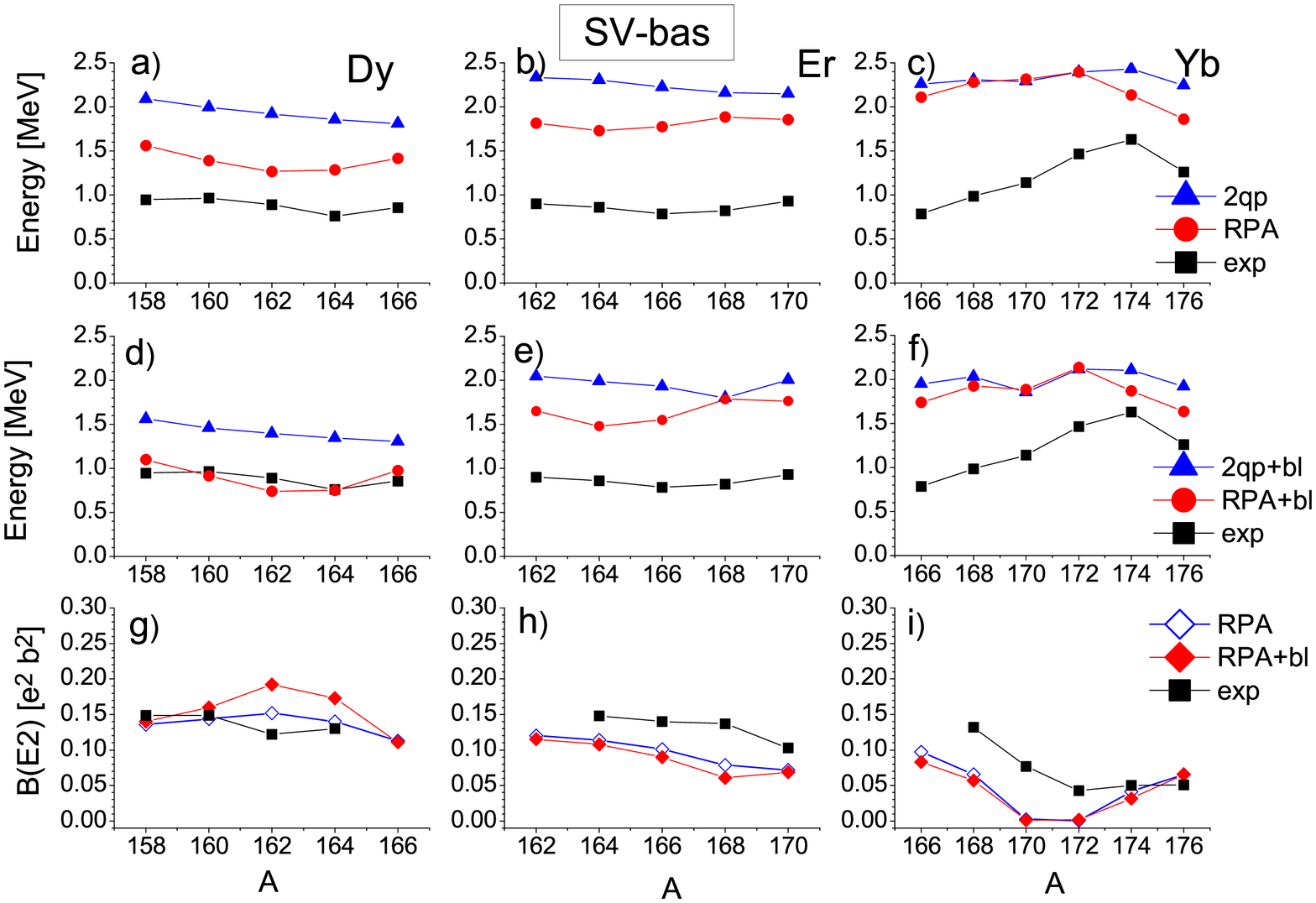}
\caption{(Color online) The SV-bas results like in Fig. 5 but for Dy, Er, and Yb isotopes.}
\end{figure*}
\begin{figure*}[t]
\includegraphics[width=12.5cm]{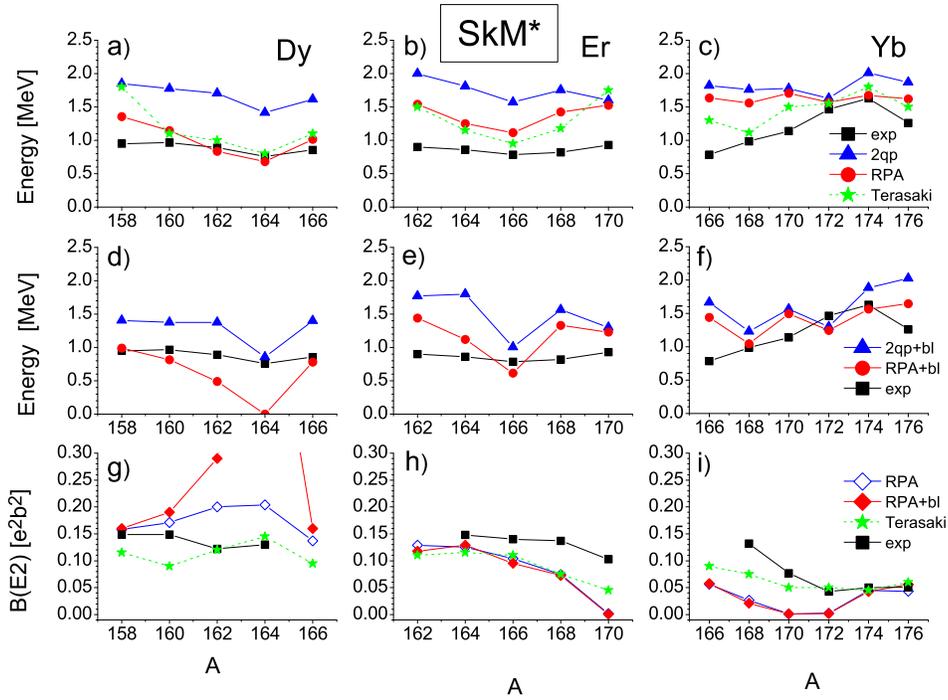}
\caption{(Color online) The same as in Fig. 6 but for  SkM$^*$.
At the plot g), the B(E2)=0.7 e$^2$b$^2$ for $^{164}$Dy is beyond
the exhibited interval.}
\end{figure*}
\begin{figure*}[t]
\includegraphics[width=12.5cm]{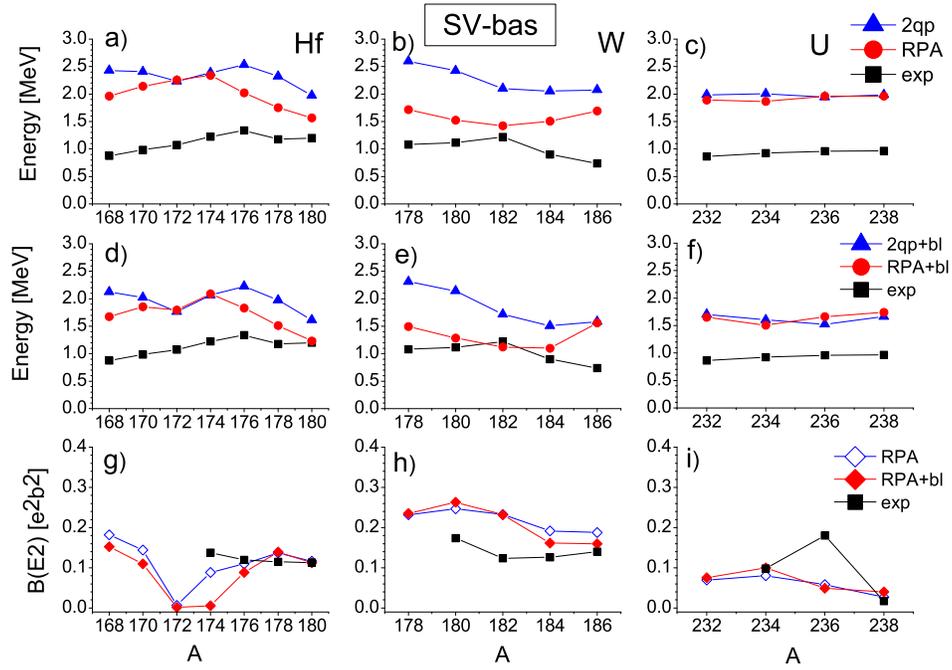}
\caption{(Color online) The SV-bas results like in Fig. 7 but
for Hf, W, and U isotopes.}
\end{figure*}
\begin{figure*}[t]
\includegraphics[width=12.5cm]{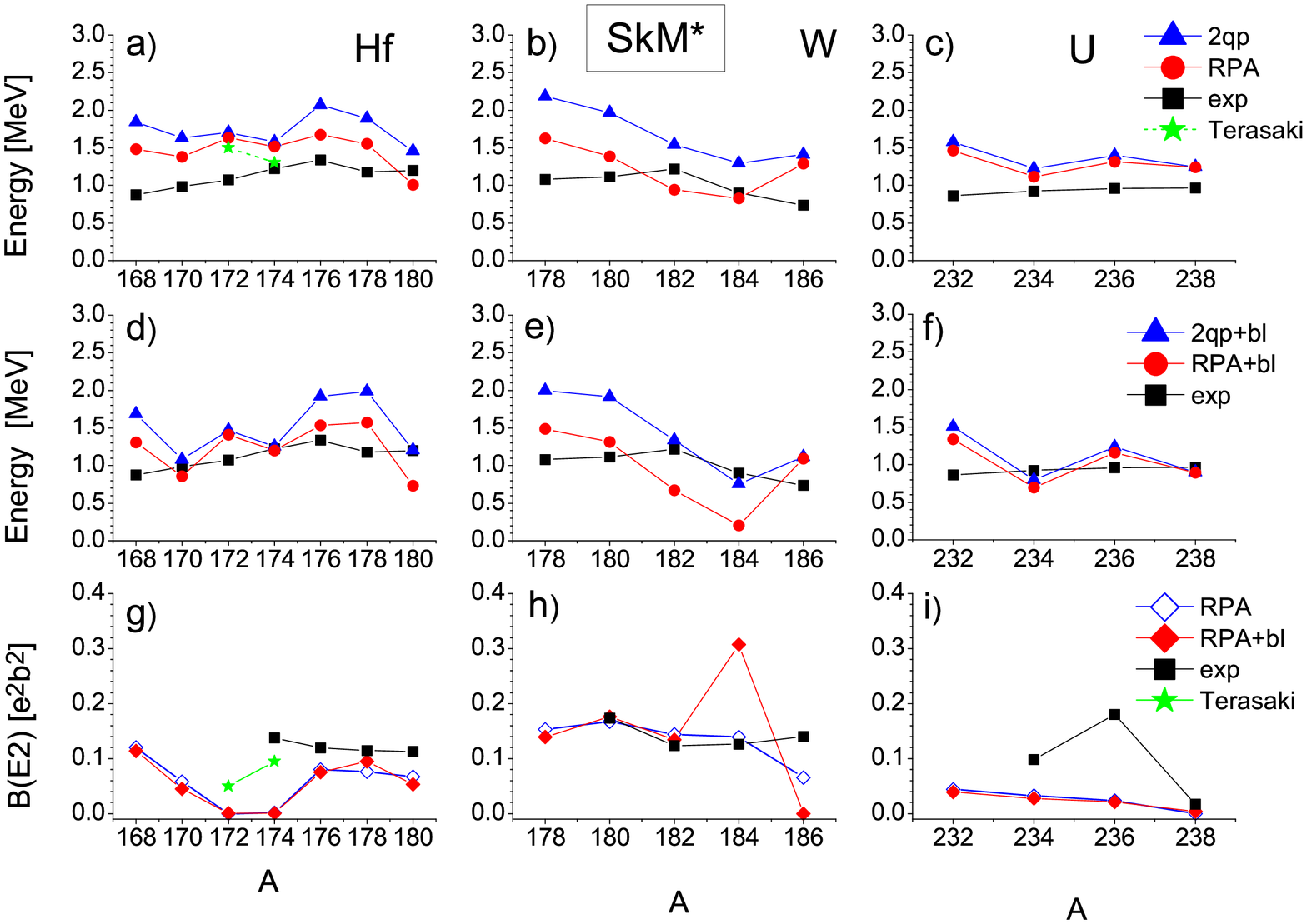}
\caption{(Color online) The same as in Fig. 6 but for the force SkM$^*$.}
\end{figure*}

Fig. 5 shows the results for Nd, Sm and Gd isotopes obtained with
SV-bas.  Calculations without PBE (plots a-c) essentially overestimate
the $2^+_{\gamma}$-energies.  The discrepancy decreases from Nd to Gd
with the growth of the collective shift $\Delta
E=\mathcal{E}_{\rm{2qp}}-E_{\rm{SRPA}}$ (the difference between the
lowest 2qp and SRPA energies). Accounting for the PBE noticeably
downshifts the 2qp energies and thus the QRPA energies (plots
d-f). The downshift reaches 0.1-0.6 MeV, depending on the isotope. As
a result, the agreement with experimental energies improves,
especially in heavy Gd isotopes. The trends of $E_{\rm{SRPA}}$ with
mass number A are approximately reproduced. The B(E2) values in Sm and
Gd with and without blocking are about the same. In Nd isotopes, the
calculated $2^+_{\gamma}$ states demonstrate a weak collectivity,
i.e. low B(E2) values.  Here the PBE worsens the agreement.  The
SkM$^*$ results in Fig. 6 for the same isotopes provide a similar
quality of description.  SRPA calculations without PBE well agree with
HFB-QRPA ones \cite{Terasaki_PRC_11}, which indicates again the
accuracy of SRPA.

Fig. 7 shows the SV-bas results for Dy, Er, and Yb isotopes.
The collectivity of calculated $2^+_{\gamma}$ states reaches
a maximum in Dy and Er isotopes.  Here we have the largest $\Delta E$ and
B(E2).  The collectivity starts to decrease in heavy Er isotopes
and almost vanishes in Yb. The PBE considerable decreases the 2qp and SRPA
energies. In Dy isotopes, this leads to a nice agreement with the
experimental energies and B(E2). In Er and Yb, the PBE noticeably
improves the description of $2^+_{\gamma}$ energies. However
$E_{\rm{RPA}}$ still remain considerably higher than $E_{\rm{exp}}$
and calculated B(E2) are accordingly underestimated.

The SkM$^*$ results for Dy-Er-Yb isotopes are given in Fig. 8. We
again observe a decrease of collectivity of $2^+_{\gamma}$ states from
Dy to Yb isotopes. However, unlike the case of light rare-earth nuclei
in Figs. 5-6, we also see a significant difference in the results
  of SV-bas and SkM$^*$. First, as compared to SV-bas results and
experimental data, the SkM$^*$ energies in Er and Yb isotopes strongly
fluctuate with A, closely following variations of 2p energies (this
feature of SkM$^*$ results was also mentioned in
\cite{Terasaki_PRC_11}).  Such fluctuations point to a small
collectivity of $2^+_{\gamma}$ states and significant contribution of
the lowest 2qp state to the structure of $2^+_{\gamma}$ state.
Furthermore, the 2qp energies are generally smaller for SkM$^*$ than
for SV-bas, which results in a better average description of
$E_{\rm{exp}}$ in Er and Yb with SkM$^*$. The PBE gives here larger
changes than for Nd-Sm-Gd isotopes.  In particular, it leads to a huge
decrease of $2^+_{\gamma}$-energy in $^{164}$Dy (like in
\cite{Terasaki_PRC_11}). This state becomes extremely collective (see
a huge overestimation of experimental B(E2)). It is unlikely that
  it can be described within a familiar QRPA and needs a more
involved prescription taking into account large ground state
correlations \cite{Hara_gsc,Vor_gsc,Klu08}. The SRPA results agree
with HFB-QRPA ones \cite{Terasaki_PRC_11} for Er-Yb but not for Dy,
especially in the exceptional case of $^{164}$Dy.

Figs. 9-10 show the results for heavy rare-earth Hf-W and actinide U
isotopes. For both forces, the collectivity of $2^+_{\gamma}$ states
increases from Hf to W and decreases in U. Moreover, both forces give
rather similar trends of $E_{\rm{SRPA}}$ with A, though deviating from
the experimental ones.  The PBE considerably downshifts the 2qp and
SRPA energies and thus in general improves their description. In average,
SkM$^*$ energies are closer to $E_{\rm{exp}}$ than SV-bas ones but
give more fuzzy A-dependence, especially with PBE. In U isotopes, the
description of the spectra with SkM$^*$ is much better than with
SV-bas, which again is explained by lower 2qp energies in
SkM$^*$. The description of B(E2) is acceptable in heavy Hf isotopes
for both SV-bas and SkM$^*$. With exception of $^{184}$W, the PBE
does not affect the description of B(E2).

Altogether, the results From Figs. 5-10 allow to do the following
conclusions: i) In rare-earth and actinide regions, there are
pronounced isotopic domains with low and high collectivity of
$2^+_{\gamma}$ states.  ii) The best agreement with the experimental
data is obtained for Dy (except for $^{164}$Dy) and W isotopes,
i.e. for the most collective $2^+_{\gamma}$ states characterized by
large $\Delta E$ and B(E2) values.  iii) The PBE essentially
downshifts 2qp and QRPA energies, thus leading to a better agreement
with experiment. The value of the downshift is comparable with the
collective shift $\Delta$E of QRPA and much larger than the
experimental errors \cite{bnl_exp}.  This indicates that the PBE plays
a non-negligible role for energies of low lying states. At the same
time, the blocking also can have a small effect on the B(E2)
values. Note that the results iii) should be checked within a truly
self-consistent PBE-QRPA approach yet to be developed.

The above conclusions are supported by both SV-bas and SkM$^*$. These
two forces give similar results in light rare-earth nuclei but deviate
in heavier nuclei. In SV-bas, the $E_{\rm{SRPA}}$ vary less with
system size $A$ but are usually larger than $E_{\rm{exp}}$. In
SkM$^*$, the variation of $E_{\rm{SRPA}}$ is stronger but this force
gives lower 2qp and SRPA energies and thus better describes
$E_{\rm{exp}}$, e.g. in U isotopes.  The differences are partly caused
by a weaker pairing in SkM$^*$ (the gaps in SkM$^*$ are in average
30-50$\%$ smaller than in SV-bas). The latter in turn can follow from
different level densities of SV-bas and SkM$^*$ s-p spectra.

It is also useful to inspect the r.m.s. deviations of the
calculated results from the experimental data,
\begin{equation}
 \sigma_b= \sqrt{
 \frac{
 \sum_{i=1}^{\mathcal{N}_b}
 (b^{\rm{cal}}_i-b^{\rm{exp}}_i)^2}
 {\mathcal{N}_b}} ,
\end{equation}
where $b^{\rm{cal}}_i$ and $b^{\rm{exp}}_i$ are calculated and experimental
values, $\mathcal{N}_b$ is the number of involved nuclei.
The deviations for the  QRPA energies ($\sigma_E$) and B(E2)-values ($\sigma_B$)
are presented in Table \ref{tab:diff}. The cases with and without PBE are estimated.
In the lower part of the Table, the SkM* SRPA deviations (without blocking) are compared
with those of Ref. \cite{Terasaki_PRC_11}
(manually obtained from the figures of \cite{Terasaki_PRC_11}).

Table \ref{tab:diff} confirms that inclusion of PBE significantly
improves description of $2^+_{\gamma}$-energies but somewhat worsens
reproduction of B(E2). This takes place for both SV-bas and SkM*.  In
agreement with previous findings, the SkM* noticeably better describes
the energies than SV-bas.  As compared to \cite{Terasaki_PRC_11}, SRPA
demonstrates the better (similar) performance for
$2^+_{\gamma}$-energies for the cases with (without) PBE. However SRPA
results are generally worse for B(E2). Perhaps the latter is caused by
the impact of the pp-channel which is included in
\cite{Terasaki_PRC_11} but skipped in SRPA.

Following Table \ref{tab:diff}, the performance of both SRPA and HFB+QRPA
\cite{Terasaki_PRC_11} is generally not good. The deviations $\sigma_{E,B}$ are large.
This calls for further improvement of the description, e.g. for inclusion of
the coupling to complex configurations (CCC). The calculated QRPA energies of
$2^+_{\gamma}$ states mostly overestimate the experimental values. Thus we still
have a window for CCC which, being a sort of additional correlations, can in
some cases downshift the energies of the lowest excited states.

\begin{table}
\caption{\label{tab:diff}Deviations between the calculated and experimental values of
$2^+_{\gamma}$-energies ($\sigma_E$)and B(E2)-strengths ($\sigma_B$). $\mathcal{N}_{E,B}$
is the number of the involved nuclei. The SRPA deviations
are compared  with ones from \cite{Terasaki_PRC_11}.}
\begin{tabular}{|c|c|c|c|c|c|c|c|}
\hline
  & Skyrme & $\mathcal{N}_E$ & \multicolumn{2}{c|}{$\sigma_E$ [MeV]}
 & $\mathcal{N}_B$ & \multicolumn{2}{c|}{$\sigma_B [e^2 b^2]$} \\
& force && no PBE & PBE &  & no PBE & PBE \\
\hline
     & SV-bas  & 40 & 0.87 & 0.62 & 31 & 0.046 & 0.056 \\
SRPA & SkM*    & 40 & 0.52 & 0.40$^{*)}$ & 31 & 0.059 & 0.075$^{*)}$ \\
     & SkM*    & 24 & 0.52 & 0.44$^{*)}$ & 18 & 0.061 & 0.072$^{*)}$ \\
\hline
Ref.\cite{Terasaki_PRC_11} & SkM* & 24 & 0.49 &  & 18 & 0.034 &  \\
\hline
\end{tabular}
\\
$^{*)}$  \footnotesize{In SkM*  SRPA(PBE) estimation for $\sigma_{\scriptsize{E,B}}$,
the anomalous nucleus $^{164}$Dy is omitted ($\mathcal{N}_E$=39(23) and
$\mathcal{N}_B$=30(17)).}
\end{table}

Note also that the description of $2^+_{\gamma}$ states depends on a
fragile balance of many factors (optimal s-p scheme, deformation,
pairing with PBE and pp-channel, CCC with the corrections from the
Pauli principle, etc) with comparable impacts. Moreover, these
ingredients have opposite effects which partly compensate each other
(e.g. the corrections from the Pauli principle may suppress the impact
of CCC \cite{Sol_Shir}).  Then, adding one of the factors, while
ignoring its balance by others, may even worsen the description. In
this connection, it would be premature to state, for example, that the
performance of SV-bas for $2^+_{\gamma}$ states is worse than of
SkM*. Also it would be wrong to state that if the effect of the
particular factor is comparable with the dependence on the Skyrme
parametrization, then this factor should be skipped.  The final
conclusions can be done only after collecting all the relevant factors
which can affect the result.

\begin{table*}
\caption{\label{tab:poles} Features of the lowest (after blocking) 2qp (i,j) and
corresponding $\lambda\mu\nu=221$  QRPA states in rare-earth nuclei,
calculated with SV-bas and SkM$^*$ forces. The table includes: the
notation $qq[Nn_z\Lambda]_i [Nn_z\Lambda]_j$ of 2qp state in Nilsson quantum
numbers; location of the s-p levels $i$ and $j$ relative to the Fermi
(F) level; the quadrupole 2qp matrix element
$f^{22}_{ij}=\langle ij|r^2 Y_{22}|0\rangle$; the 2qp energy
$\epsilon^q_{ij}$ (\ref{eps_ij}) and collective shift $\Delta
E=\epsilon^q_{ij}-E_{221}$, calculated without the blocking; the 2qp
energy $\mathcal{E}^q_{\rm{bl}}(ij)$ (\ref{37}) and collective shift
$\Delta E_{\rm{bl}}=\mathcal{E}^q_{\rm{bl}}(ij)-E_{221}$, calculated
with the blocking; the blocking correction $\Delta
\mathcal{E}^q_{\rm{bl}}=\epsilon^q_{ij}-\mathcal{E}^q_{\rm{bl}}(ij)$.  See
text for more detail.}
\begin{tabular}{|c|c|c|c|c|c|c|c|c|c|}
\hline
Nucleus & Force & $qq[Nn_z\Lambda]_i [Nn_z\Lambda]_j$ & F-location & $f^{22}_{ij}$ &
$\epsilon^q_{ij}$ & $\Delta E$ & $\mathcal{E}^q_{\rm{bl}}(ij)$ &
$\Delta E_{\rm{bl}}$ & $\Delta \mathcal{E}^q_{\rm{bl}}$ \\
&&&& [fm$^4$] & [MeV] & [MeV] &  [MeV] &  [MeV] &  [MeV] \\
\hline
 $^{154}_{\;\;62}$Sm$_{92}$ & SV-bas & pp[413]$\downarrow$[411]$\downarrow$ & F, F+3  & -4.43 & 2.57 & 0.46 & 2.34 & 0.38 & 0.23 \\
            & SkM$^*$ & pp[411]$\downarrow$[411]$\uparrow$ & F+3, F+1   & 4.98 & 2.45 & 0.34 & 2.37 & 0.31 & 0.07 \\
 $^{162}_{\;\;66}$Dy$_{96}$ & SV-bas & pp[411]$\downarrow$[411]$\uparrow$ & F+1, F  & 6.58 & 1.92 & 0.65 & 1.39 & 0.65 & 0.53 \\
            & SkM$^*$ & pp[413]$\downarrow$[411]$\downarrow$ & F, F+1   & -5.78 & 1.71 & 0.87 & 1.37 & 0.88 & 0.33 \\
$^{164}_{\;\;66}$Dy$_{98}$ & SV-bas & pp[411]$\downarrow$[411]$\uparrow$ & F+1, F  & 6.59 & 1.86 & 0.57 & 1.34 & 0.59 & 0.51 \\
            & SkM$^*$ & nn[523]$\downarrow$[521]$\downarrow$ & F, F+1   & 5.98 & 1.42 & 0.56 & 0.86 & 0.86 & 0.56 \\
 $^{172}_{\;\;70}$Yb$_{102}$  & SV-bas & nn[512]$\uparrow$[521]$\downarrow$ & F+1, F-1 & 0.37 & 2.40 & 0.003 & 2.12 & -0.02 & 0.28 \\
            & SkM$^*$ & nn[512]$\uparrow$[521]$\downarrow$ & F+1, F-1   & 0.086 & 1.63 & 0.06 & 1.30 & 0.06 & 0.33 \\
 $^{174}_{\;\;72}$Hf$_{102}$ & SV-bas & nn[512]$\uparrow$[521]$\downarrow$ & F+1, F-1 & 0.37 & 2.39 & -0.02 & 2.07 & 0.05 & 0.32 \\
            & SkM$^*$ & nn[512]$\uparrow$[521]$\downarrow$ & F+1, F-1   & 0.19 & 1.58 & 0.06 & 1.26 & 0.06 & 0.33 \\
 $^{176}_{\;\;72}$Hf$_{104}$ & SV-bas & nn[512]$\uparrow$[510]$\uparrow$ & F, F-2    & -8.17 & 2.48 & 0.47 & 2.14 & 0.34 & 0.33 \\
            & SkM$^*$ & nn[512]$\uparrow$[510]$\uparrow$ & F, F-2       & -8.48 & 2.53 & 0.51 & 2.23  & 0.39 & 0.31 \\
 $^{182}_{\;\;74}$W$_{108}$  & SV-bas & nn[510]$\uparrow$[512]$\downarrow$ & F+1, F+2 & 8.82 & 2.10 & 0.68 & 1.72 & 0.59 & 0.39 \\
            & SkM$^*$ &  nn[510]$\uparrow$[512]$\downarrow$ & F+1, F+2 & 7.98 & 1.54 & 0.60 & 1.34 & 0.67 & 0.21 \\
\hline
\end{tabular}
\end{table*}

\subsection{Discussion}

In this subsection, we analyze the above results and compare them with
earlier studies \cite{Terasaki_PRC_11,Sol_Grig,Sol_Shir,Sol_Sush_Shir}.

First of all, it is worth to explore the origin of domains with low
and high collectivity of $2^+_{\gamma}$ states. The low-collectivity
domains include most of Nd, Er, Yb, Hf, and U isotopes. High collectivity
exists in Sm, Gd, Dy, and W isotopes. Table \ref{tab:poles} shows that the
appearance of such domains is determined by the structure of the first
2qp states which, in turn, results in different absolute values of the
matrix element $f^{22}_{ij}=\langle ij|r^2Y_{22}|0\rangle$
for the doorway operator $r^2 Y_{22}$.
These 2qp states are built from the levels close to the the Fermi level.
High collectivity (pertinent to $^{154}$Sm, $^{162,164}$Dy,
$^{176}$Hf, and $^{182}$W) takes place if the state is
characterized by a large value of $|f^{22}_{ij}|$. Instead, if $|f^{22}_{ij}|$ is
small, then we get non-collective $2^+_{\gamma}$ states ($^{172}$Yb
and $^{174}$Hf).  The magnitude of $|f^{22}_{ij}|$ is determined by
Nilsson selection rules for E2(K=2) transitions in axial nuclei
\cite{Nilsson65,Sol76}. The rules read
\begin{equation}
  \Delta K = 2,
  \;\;
  \Delta N =0, \pm 2,
  \;\;
  \Delta n_z=0,
  \;\;
  \Delta \Lambda =2 ,
\label{eq:select}
\end{equation}
where $N$ is the principle quantum shell number, $n_z$ is the fraction
of $N$ along the z-axis, $\Lambda$ is the orbital momentum projection
onto z-axis. All the 2qp states in Table \ref{tab:poles} fulfill the rules
(\ref{eq:select}) for $K$ and $N$ but not for $n_z$ and $\Lambda$.
Table \ref{tab:poles} shows that the rule $\Delta n_z=0$ is decisive. The
2qp states which keep this rule ($^{154}$Sm, $^{162,164}$Dy,
$^{176}$Hf, $^{182}$W) exhibit $|f^{22}_{ij}|$-values of one order of
magnitude larger than states violating the rule ($^{172}$Yb and $^{174}$Hf).
This effect is especially spectacular for neighboring isotopes
$^{174}$Hf - $^{176}$Hf.  The rule $\Delta \Lambda =2$ is not so
crucial. However, matrix elements are additionally increased if this rule
is obeyed  ($^{176}$Hf, $^{182}$W).

Table \ref{tab:poles} obviously suggests that just the strength
$|f^{22}_{ij}|$ of the first 2qp state is decisive for the
collectivity of the QRPA $2^+_{\gamma}$ state and formation of the
domains with low and high collectivity.  This finding can be
corroborated within a simple two-pole model given in Appendix
C. Following this model, the collectivity of the lowest QRPA states is
mainly determined by the ratio between the strengths of the first
($\nu$ =1) and second ($\nu$ =2) 2qp states where the second state
simulates a cumulative effect of all 2qp states with
$\nu>$1. Depending on this ratio, different scenarios can take place:
high-collective limit, intermediate case and low-collective limit. In
the last case, the first QRPA energy can lie even a bit above the
first 2qp state, which happens, e.g., in our calculations for Yb
isotopes.

Altogether, we get a simple recipe for predicting the collectivity of
the first QRPA state: it suffices to inspect the Nilsson selection
rules (\ref{eq:select}) for the lowest 2qp state, first of all $\Delta
n_z=0$. Note that, unlike s-p spectra, the s-p wave functions and thus
the values $|f^{22}_{ij}|$ only slightly depend on the Skyrme
parametrization \cite{Nest04}, which makes the proposed recipe quite
reliable. As seen from Table 1, SV-bas and SkM$^*$ sometimes give
different lowest 2qp states. Nonetheless, the correlation between
$\Delta n_z=0$ rule and collectivity of QRPA $2^+_{\gamma}$-states
applies in all considered cases.

The nucleus $^{164}$Dy computed with SkM$^*$ shows a remarkable
sequence of four strong ($|f^{22}_{ij}|$=5.8-9.2 fm$^4$) 2qp states
which are  located with PBE at 0.86 - 1.96 MeV.  The cumulative impact
of these states delivers a dramatic effect: a break-down of
RPA. Without PBE, these four 2qp states lie at a higher energy 1.42--2.15
MeV and do not lead to the instability.  For comparison, SV-bas
gives in $^{164}$Dy only three strong ($|f^{22}_{ij}|$=5.4-6.6 fm$^4$)
2qp states and they are located at a higher energy 1.35-1.65 MeV. This
gives a collective $2^+_{\gamma}$-state still within QRPA.  Altogether,
this discussion shows that some QRPA results for low lying states can be
quite sensitive to the Skyrme force.

Table \ref{tab:poles} shows that the values of collective shifts $\Delta E$ (up to
0.9 MeV) and blocking induced shifts $\Delta \mathcal{E}_{\rm{bl}}$
(up to 0.6 MeV) are comparable.  Thus the PBE has a
non-negligible effect in the present calculations.

The results exhibited in Figs. 5-10 indicate that the present Skyrme
QRPA description of $2^+_{\gamma}$ states is not yet fully
satisfactory.  Though we get rather good agreement with experimental
data for collective $2^+_{\gamma}$ states in Gd, Dy, and W isotopes,
collectivity is generally underestimated in other isotopic chains
(which is seen from too high SRPA energies and sizable low
B(E2)-values).  Perhaps the latter cases require a coupling to complex
configurations, which might affect both the $2^+_{\gamma}$-energies
and  $B(E2)$-values. In this respect,
our calculations indicate regions where CCC is needed. In the previous
Skyrme QRPA study \cite{Terasaki_PRC_11}, the need for CCC was also
pointed out.  In nuclei like $^{164}$Dy, an approach taking into
account large ground state correlations is necessary
\cite{Hara_gsc,Vor_gsc}.

As seen in Figs. 5-10, the performances of our and previous
\cite{Terasaki_PRC_11} systematic Skyrme QRPA calculations (without the
PBE) are rather similar.
Although these calculations exploit different prescriptions, HFB + exact QRPA
in \cite{Terasaki_PRC_11} and BCS+PBE + separable QRPA in the present
study, they provide a remarkably similar description of QRPA energies
of $2^+_{\gamma}$ states. The results \cite{Terasaki_PRC_11} are
somewhat better for B(E2)-values, though the difference is not
crucial.

Since SRPA operates with the residual interaction in a
separable form, it can be directly compared with
schematic separable  QRPA approaches, e.g. with QPM which is widely and
successfully used in nuclear spectroscopy \cite{Sol76}. The QPM
proposes some simple relations for the strength constants of the
residual interaction which might be useful for a rough evaluation of
the SRPA strength constants. This analysis is done in the
Appendix C. It is shown that the mixed isoscalar-isovector  interaction might be
essential in Skyrme QRPA. If this interaction is not properly balanced,
it can weaken a general isoscalar effect of the residual interaction
and thus make $2^+_{\gamma}$ states less collective (which might be
relevant for Nd, Yb, Hf, U isotopes).

\section{Summary}

We have performed a systematic study of the lowest
$\gamma$-vibrational $K^{\pi}=2^+$ states in axially
deformed even-even  rare-earth and actinide
nuclei within a self-consistent (except for the pairing part) separable
random-phase-approximation (SRPA)
\cite{Ne06}.  Nine isotopic chains involving 41 nuclei were explored.
The excitation energies and B(E2)-values of $2^+_{\gamma}$ states were
computed and analyzed. The Skyrme forces SV-bas \cite{SV} and
SkM$^*$ \cite{SkMs} were used. The force SV-bas was chosen as
providing a good description of ground state deformations and
isoscalar giant quadrupole resonance (ISGQR).
SkM$^*$ was used as a force with the best performance
in the previous systematic study of $2^+_{\gamma}$ states \cite{Terasaki_PRC_11},
performed within the exact (not factorized) Skyrme HFB+QRPA.
The accuracy of SRPA was confirmed  by comparison with calculations
within exact BCS+QRPA \cite{repko} and BCS+QRPA \cite{Terasaki_PRC_11}.

Our study undertakes some important steps which were not realized
earlier \cite{Terasaki_PRC_11}.  Some essential points concerning the
pairing contribution, systematics of $2^+_{\gamma}$ states and
explanation of the results were scrutinized.

First, we have investigated a possible impact of the pairing blocking
effect (PBE) on the properties of $2^+_{\gamma}$ states.  Thereby we
use in "ad hoc" manner from the PBE only the correction of 2qp
energies while the 2qp wave functions remain the same as in the BCS
ground state. This scheme has significant advantages: it incorporates
the most essential energy correction from PBE but maintains, at the
same time, the orthonormality of the 2qp configuration space which, in
turn, allows to apply the standard QRPA solution scheme. This blocking
scheme was applied to a few lowest two-quasiparticle (2qp)
configurations whose corrected energies were then used in SRPA
calculations.  Within this scheme, the PBE significantly downshifts
the SRPA energies of $2^+_{\gamma}$ states and thus improves agreement
with the experimental spectra. At the same time, PBE rather slightly
affects collectivity of the states, expressed in terms of collective
shifts and transition probabilities B(E2).  It is to be noted, that
our present handling of the PBE is very preliminary and should be
further checked in fully developed self-consistent QRPA with PBE.
To the best of our knowledge, such methods are still absent. Then
our study can be viewed as a first step which highlights the problem
and calls for a further self-consistent exploration. Note  also that
the PBE-QRPA scheme is certainly not the only way to improve the description of
$2^+_{\gamma}$ states. Various many-body techniques that go beyond the plain QRPA,
first of all the coupling to complex configuration, can be decisive here.

As the next novel aspect of our study, we have singled out domains
of nuclei with a low and high collectivity of $2^+_{\gamma}$
states. It was shown that collectivity is mostly determined by the
structure of the lowest 2qp state constituting the first SRPA 2qp
state.  The effect was explained in terms of the Nilsson selection
rule $\Delta n_z$=0, which delivers a simple recipe to predict the
$2^+_{\gamma}$-collectivity without performing QRPA calculations. Some
results and SRPA characteristics were compared with those from the
schematic Quasiparticle-Phonon Model (QPM) \cite{Sol76} which was
successfully used for a long time in nuclear spectroscopy.

It was found that the forces SV-bas and SkM$^*$ perform similarly in
the description of $2^+_{\gamma}$ states for light rare-earth nuclei
but deviate in heavier nuclei. The latter is mainly explained by the
fact that SkM$^*$ delivers a weaker pairing gap and thus lower 2qp
energies, than SV-bas.  SV-bas delivers less fuzzy trends of energies
and B(E2) values and well describes Dy isotopes but fails in U
isotopes.  SkM$^*$ is better in U isotopes but its results fluctuate
more with the mass number. Moreover, SV-bas has an important advantage
over SkM$^*$: it well describes quadrupole equilibrium deformations
and energy centroids of ISGQR. Thus SV-bas allows to get a consistent
description of $2^+_{\gamma}$ states and ISGQR.

In general our study shows that, despite all the progress,
available fully or partly self-consistent QRPA schemes
are still not accurate enough for a satisfactory description of
$2^+_{\gamma}$ states throughout medium and heavy axially deformed nuclei.
This holds for both our results and previous ones \cite{Terasaki_PRC_11}.
Some essential factors should be still added or improved. The proper calculation scheme
should fulfill at least the following requirements: a) accurate description
of the s-p spectra and equilibrium deformation, b) treatment of pairing (BCS or HFB)
with PBE, c) self-consistent residual QRPA interaction
with both ph- and pp-channels and consistently incorporated PBE,
d) simultaneous description of other quadrupole excitations (ISGQR),
e) systematic description involving nuclei from various mass regions
and domains with a low and high collectivity,
f) the coupling to complex configuration  (with the proper inclusion
of the Pauli principle).  Some of
these points will be a subject of our next studies.

\section*{Acknowledgments}
The work was partly supported by the DFG grant RE 322/14-1, Heisenberg-Landau
(Germany-BLTP JINR), and Votruba-Blokhintsev (Czech Republic-BLTP JINR)
grants. The BMBF support under the contracts 05P12RFFTG (P.-G.R.)  and
05P12ODDUE (W.K.) is appreciated. J.K. is grateful for the support
of the Czech Science Foundation (P203-13-07117S). We thank J. Terasaki,
A.V. Sushkov and A. P. Severyukhin for useful discussions.

\appendix
\section{Pairing cut-off weight and pairing matrix elements}

To simulate the effect of a finite range pairing force, the
pairing-active space for each isospin $q$ is limited by using a
smooth energy-dependent cut-off (see e.g. \cite{BCS50,Bon85})
\begin{equation}
 f^q_k
 =
 \frac{1}
      {1+\rm{exp}[\frac{\tilde{e}^q_k-\lambda_{q}-\Delta E_{q}}{\eta_{q}}]}
\end{equation}
in the sums in Eqs. (\ref{34}), (\ref{35}), (\ref{38}), and
(\ref{30}).  The cut-off parameters $\Delta E_{q}$ and
$\eta_{q}=\Delta E_{q}/10$ are chosen self-adjusting to the actual level
density in the vicinity of the Fermi energy, see \cite{Ben00} for details.

For the $\delta$-force pairing interaction (\ref{5}), the
anti-symmetrized pairing matrix elements read
\begin{eqnarray}\label{26}
  V_{i \bar{i} j \bar{j}}^{(\rm{pair,q})}
  &=&
  \langle i \bar{i}|V^q_{\rm{pair}} ({\mathbf{r}}, {\mathbf{r}}') |j \bar{j}\rangle_q
\\
  &=&
  \int d^3r \int d{\mathbf{r}}'
  \:\Phi^+_i({\mathbf{r}})  \Phi^+_{\bar{i}}({\mathbf{r}}')\: V_q
 \; \delta({\mathbf{r}} - {\mathbf{r}}')
\nonumber \\
  &&
  \cdot [\Phi_j({\mathbf{r}}) \Phi_{\bar{j}}({\mathbf{r}}')
  - \Phi_j({\mathbf{r}}') \Phi_{\bar{j}}({\mathbf{r}})]
\nonumber \\
  &=&
  V_q
  \int d^3r
  [\left( \Phi^+_i(\mathbf{r}) \cdot \Phi_j(\mathbf{r})\right) \:
  \left( \Phi^+_{\bar{i}}(\mathbf{r}) \cdot \Phi_{\bar{j}}(\mathbf{r}) \right)
\nonumber \\
  &&
  -\left( \Phi^+_i(\mathbf{r}) \cdot \Phi_{\bar{j}}(\mathbf{r}) \right)
   \left( \Phi^+_{\bar{i}}(\mathbf{r}) \cdot \Phi_j(\mathbf{r}) \right)]
\nonumber
\end{eqnarray}
where
\begin{equation}
  \Phi_i(\mathbf{r})
  =
  \left(
  \begin{array}{rl}
  R^{(+)}_i(\rho,z)\:\: e^{i (K_i-\frac{1}{2}) \vartheta } \\
  R^{(-)}_i(\rho,z)\:\: e^{i (K_i+\frac{1}{2}) \vartheta }
  \end{array}
  \right) ,
\end{equation}
\begin{equation}
  \Phi_{\bar{i}}(\mathbf{r})
  =
  \left(
  \begin{array}{rl}
  - R^{(-)}_i(\rho,z)\:\: e^{-i (K_i+\frac{1}{2}) \vartheta } \\
  R^{(+)}_i(\rho,z)\:\: e^{-i (K_i-\frac{1}{2}) \vartheta }
  \end{array}
  \right)
\label{27}
\end{equation}
are spinor s-p. wave functions in cylindrical coordinates
$(\rho,z,\vartheta)$ and
$\left(\Phi^+_i({\mathbf{r}})\cdot\Phi_j({\mathbf{r}})\right)$ are
scalar products. Denoting the first (Hartree) and second (exchange)
terms in the last line of (\ref{26}) as
$V_{i\bar{i}j\bar{j}}^{(\rm{pair-H, q})}$ and
$V_{i\bar{i}j\bar{j}}^{(\rm{pair-ex, q})}$, we obtain
\begin{eqnarray} \label{28b}
  &&
  V_{i \bar{i} j \bar{j}}^{(\rm{pair-H, q})}
  =
  2 \pi \: V_q \int_{0}^{\infty} d\rho \: \int_{-\infty}^{\infty} dz \: \rho
\\
  &&
  \left[ 2 R_i^{(+)} R_j^{(+)} R_i^{(-)} R_j^{(-)} +
  ( R^{(-)}_i R^{(-)}_j )^2 + ( R^{(+)}_i R^{(+)}_j )^2 \right]
  \;,
\nonumber
\end{eqnarray}
\begin{eqnarray}\label{28c}
  &&
  V_{i \bar{i} j \bar{j}}^{(\rm{pair-ex, q})}
  =
  2 \pi \: V_q \int_{0}^{\infty} d\rho \: \int_{-\infty}^{\infty} dz \: \rho
\\
&&
  \left[ - 2 R_i^{(+)} R_j^{(-)} R_i^{(-)} R_j^{(+)} +
   ( R^{(-)}_i R^{(-)}_j )^2 + ( R^{(+)}_i R^{(+)}_j )^2 \right]
 \nonumber
\end{eqnarray}
and finally
\begin{eqnarray}\label{28}
  V_{i \bar{i} j \bar{j}}^{(\rm{pair, q})}
  &=&
  V_{i \bar{i} j \bar{j}}^{(\rm{pair-H, q})}
  + V_{i \bar{i} j \bar{j}}^{(\rm{pair-ex, q})}
\\
  &=&
  2 \pi \: V_q \int_{0}^{\infty} d\rho \: \int_{-\infty}^{\infty} dz \: \rho
\nonumber
\\
  &&
  \left[(R_i^{(+)})^2 + (R_i^{(-)})^2] \; [(R_j^{(+)})^2 +
    (R_j^{(-)})^2] \right]
  \;.
\nonumber
\end{eqnarray}

\section{Basic SRPA equations}

The self-consistent derivation \cite{Ne06,Ne_ar05} yields the SRPA Hamiltonian
\begin{equation}\label{Ham}
    \hat{H} = \sum_q \hat{h}^q_{\mathrm{HF+BCS}} + \hat{V}_{\mathrm{res}}
\end{equation}
where
\begin{equation}\label{eq:HFBmf}
    \hat{h}^q_{\mathrm{HFB}} = \int d{\mathbf{r}} \sum_{\alpha, \alpha'}
    \left[
    \frac{\delta E}{\delta J^q_{\alpha} ({\mathbf{r}})} \hat{J}_q^{\alpha}({\mathbf{r}})
    \right]
\end{equation}
is the mean field and pairing contribution and
\begin{eqnarray}
\label{res_int}
    \hat{V}_{\rm{res}} &=&
    \frac{1}{2}
    \sum_{qq'} \sum_{m, m =1}^{M}
[\kappa_{qm, q'm'} \hat{X}_{qm} \hat{X}_{q'm'}
    \\
    &+& \eta_{qm, q'm'} \hat{Y}_{qm} \hat{Y}_{q'm'}]
    \nonumber
\end{eqnarray}
is the separable residual interaction with one-body operators
\begin{eqnarray}\label{Xop}
    \hat{X}_{qm}&=&\sum_{q'} \hat{X}^{q'}_{qm} = i \sum_{q'} \sum_{\alpha, \alpha'}
    \int d{\mathbf{r}}
\\
 &&
 \left[
    \frac
    {\delta^2 E}{\delta J^{q'}_{\alpha'} ({\mathbf{r'}})
    \delta J^{q}_{\alpha} ({\mathbf{r})}}
    \right]
    \langle
    \left[\hat{P}_{qm}, \hat{J}^{q}_{\alpha}(\mathbf{r})
    \right]
    \rangle
    \hat{J}^{q'}_{\alpha'} (\mathbf{r'}) \; ,
    \nonumber\\
 \label{Yop}
     \hat{Y}_{qm}&=&\sum_{q'} \hat{Y}^{q'}_{qm} = i \sum_{q'} \sum_{\alpha, \alpha'}
    \int d{\mathbf{r}}
\\
 &&
    \left[
    \frac{\delta^2 E}{\delta J^{q'}_{\alpha'} ({\mathbf{r'}})
    \delta J^q_{\alpha} ({\mathbf{r})}}
 \right]
   \langle \left[ \hat{Q}_{qm}, \hat{J}^{q}_{\alpha}(\mathbf{r}) \right]\rangle
    \hat{J}^{q'}_{\alpha'} (\mathbf{r'})
\nonumber
\end{eqnarray}
and inverse strength matrices
\begin{eqnarray}
\label{eq:kappa_eta}
  \kappa_{qm q'm'}^{-1 } &=&
  - i \langle [\hat{P}_{qm},{\hat X}_{q'm'}] \rangle ,
  \\
  \eta_{qm q'm'}^{-1 }
  &=& -i
  \langle [\hat{Q}_{qm},{\hat Y}_{q'm'}] \rangle \; .
\end{eqnarray}
Here $\alpha = \rho, \tau, \mathbf{J}, \chi, \mathbf{j}, \mathbf{s},
\mathbf{T}$ enumerates densities $J^q_{\alpha}$ and their operators
$\hat J^q_{\alpha}$ while $m$ marks time-even $\hat{Q}_{qm}$ and
time-odd $\hat{P}_{qm}=i[\hat H,\hat{Q}_{qm}]$ Hermitian input
(doorway) operators. The number $M$ of separable terms in
(\ref{res_int}) is determined by the number of the input operators
$\hat{Q}_{qm}$ chosen from physical arguments \cite{Ne06,Ne02}.  Usually we
have $M=$3--5. For then, the QRPA matrix has a low rank $4M$ and we
have small computational expense even for heavy deformed nuclei.

The values $\langle \left[\hat{P}_{qm}, \hat{J}^{q}_{\alpha}
  \right]\rangle$ from (\ref{Xop}) and $\langle \left[ \hat{Q}_{qm},
  \hat{J}^{q}_{\alpha} \right]\rangle$ from (\ref{Yop}) do not vanish
only for time-even and time-odd densities $\hat{J}^{q}_{\alpha}$,
respectively.  Then $\hat{X}_k$ is time-even (determined by time-even
densities) while $\hat{Y}_k$ is time-odd (determined by time-odd
densities). The SRPA residual interaction (\ref{res_int}) includes
contributions from variations of both time-odd and time-even
densities.

Following (\ref{eq:HFBmf}), (\ref{Xop}) and (\ref{Yop}),
$\hat{h}_{\mathrm{HF+BCS}}$ and $\hat{V}_{\mathrm res}$ are determined
by first and second functional derivatives of the given energy
functional. The model is self-consistent for exception of the pairing part.

The operators $\hat{Q}_{qm}$ constitute the key input for SRPA
\cite{Ne06,Ne02}. They are chosen from physical arguments, namely to
produce doorway states for particular excitations.  In present
calculations, four operators are used. The first one,
$\hat{Q}_{q1}({\mathbf{r}})=r^2Y_{22}(\theta)+ \rm{h.c.}$, generates
the quadrupole ($\lambda\mu$=22) mode of interest in the long-wave
approximation ($Y_{22}(\theta)$ is the spherical harmonic).  Usually,
already one such operator (generator) is enough for a rough description of the
spectrum. However the corresponding Tassie mode \cite{Ri80, Tassie}
is mainly of the
surface character.  So, to improve accuracy of the description,
two other generators, $\hat{Q}_{q2}({\mathbf{r}})=r^4Y_{22}(\theta)+
\rm{h.c.}$ and $\hat{Q}_{q3}({\mathbf{r}})=j_2(0.6r) Y_{22}(\theta)+
\rm{h.c.}$ (with $j_2(0.6r)$ being the spherical Bessel function), are
added. These generators result in $\hat{X}^{q'}_{qm}({\mathbf{r}})$
operators peaked more in the nuclear interior \cite{Ne06}. Finally,
the generator $\hat{Q}_{q4}({\mathbf{r}})=r^4Y_{42}(\theta)+
\rm{h.c.}$ is added to take into account the coupling between
quadrupole and hexadecapole excitations in axially deformed
nuclei. Note that these input operators do not form directly the
separable residual interaction (\ref{res_int}) but generate its
operators $\hat{X}^{q'}_{qm}({\mathbf{r}})$,
$\hat{Y}^{q'}_{qm}({\mathbf{r}})$ and strength constants $\kappa_{qm,
  q'm'}$, $\eta_{qm, q'm'}$, based on the initial Skyrme
functional. The number $M$ of input operators determines the number of
the separable terms in (\ref{res_int}). Larger $M$ brings the
separable interaction closer to the true (not factorized) one, but
makes SRPA calculations more time consuming. The four
operators which we are using here constitute a good compromise between
reliability and expense.

SRPA allows to calculate the energies $\omega_{\nu}$ and wave function
(with forward $\psi^{\nu}_{ij}$ and backward $\phi^{\nu}_{ij}$ 2qp
amplitudes) of one-phonon $\nu$-states. Besides, various strength
functions can be directly computed (without calculation of
$\nu$-states). In this study, we use for description of ISGQR
the strength function
\begin{equation} \label{sf}
 S\:_{\gamma}(E22,\;E) =  \sum_{\nu} \:
|\:\langle\nu|\: r^2 Y_{22}\:|0\rangle \:|^2
\:\xi_{\Delta}(E-E_{\nu})
\end{equation}
where $\xi_{\Delta}(E-E_{\nu}) = \Delta/[2\pi((E-E_{\nu})^2 +
(\Delta/2)^2] $ is the Lorentz weight with the averaging parameter
$\Delta$= 1 MeV.

The energy centroids for ISGQR depicted in Fig. 3 are estimated for the energy
intervals where the strength functions exceeds 20$\%$ of its maximal value.

\section{Simple two-pole RPA model}

Let's consider SRPA with one input (doorway) operator and without
time-odd contributions. Then the SRPA secular equation is reduced to
the familiar equation for the schematic separable RPA \cite{Sol76,Ri80}:
\begin{equation}\label{se1}
 \kappa^{-1}=\sum_{ij}\frac{f_{ij}^2}{\epsilon_{ij}^2-E_{\nu}^2}
\end{equation}
where $\kappa$ is the strength constant, $f_{ij}$ is the matrix
element of the residual interaction (including the pairing factors)
between the states $i$ and $j$, $\epsilon_{ij}$ is the 2qp energy, and
$E_{\nu}$ is the energy of the $\nu$-th RPA states.  This equation may
be simplified to the case of two  2qp states, yielding two poles
  in the schematic RPA equation:
\begin{equation}\label{se2}
 1
 =
 \kappa f^2[\frac{k^2}{\epsilon^2_1-E^2}+\frac{1}{\epsilon^2_2-E^2}] .
\end{equation}
Here the first pole is characterized by the 2qp energy $\epsilon_1$ and
matrix element $kf$.  The second pole (with the 2qp energy
$\epsilon_2 > \epsilon_1$
and matrix element $f$) is assumed to simulate the effect of all the
poles above the lowest one. The coefficient $k$ determines the ratio
between the matrix elements of the first and second poles. We suppose
$\kappa > 0$, which is common for low-energy isoscalar excitations
\cite{Sol76}.

Equation (\ref{se2}) is reduced to a standard  quadratic equation
\begin{equation}
E^4 + b E^2 + c = 0
\end{equation}
with
\begin{eqnarray}
 b&=&-(\epsilon_1^2+\epsilon_2^2)+\kappa f^2 (1+k^2),
 \\
c&=&\epsilon_1^2\epsilon_2^2-\kappa f^2(\epsilon_1^2+k^2\epsilon_2^2).
\end{eqnarray}
This equation allows to get useful analytical estimations for three
important cases: i) $k \gg 1$ (strong first pole, typical for Gd, Dy,
and W isotopes), ii) $k \ll 1$ (weak first pole, typical for Nd, Yb,
Hf, and U isotopes), iii) $k = 1$ (intermediate case with equal
strengths of the first and second poles).

We go through these three cases step by step:
\begin{description}
\item{i)} For the strong first pole ($k \gg 1$), we get
$(1\pm k^2) \to \pm k^2$ and so
\begin{equation}
E^2 \approx \frac{1}{2}[\epsilon_1^2+\epsilon_2^2-\kappa (fk)^2
\pm(\epsilon_1^2-\epsilon_2^2-\kappa(fk)^2)]
\end{equation}
with two solutions
\begin{equation}
E^2_+\approx \epsilon^2_1 -\kappa (fk)^2, \;\;\;
E^2_-\approx  \epsilon^2_2
\end{equation}
The solution $E_+$ gives the energy of the 1st RPA state
below the first pole which is a common case in
phenomenological QPM
\cite{Sol76,Sol_Grig,Sol_Shir}.
In our calculations, this case is met in Gd, Dy, and W isotopes.

\item{ii)} For the weak first pole ($k \ll 1$),
we get $(1\pm k^2) \to 1$ and so
\begin{eqnarray}
E^2&\approx& \frac{1}{2}[\epsilon_1^2+\epsilon_2^2-\kappa f^2 \pm(\epsilon_1^2-\epsilon_2^2+\kappa f^2)],
\\
E^2_+ &\approx& \epsilon^2_1, \;\;\;
E^2_- \approx  \epsilon^2_2 -\kappa f^2.
\end{eqnarray}
The solution $E_+$ is the energy of the 1st RPA state
close to the first pole. This energy can be both
a bit smaller or larger than $e_1$.
We have this case for Nd, Yb and Hf isotopes.

\item{iii)} If the pole strengths are equal
($k=1$), then  $(1 - k^2) \to 0, \; (1 + k^2) \to 2$
and
\begin{equation}
E^2\approx\frac{1}{2}[\epsilon_1^2+\epsilon_2^2-2\kappa f^2
\pm\sqrt{(\epsilon_1^2-\epsilon_2^2)^2+4\kappa^2 f^4}.
\end{equation}
Supposing that $(\epsilon_1^2-\epsilon_2^2)^2 \gg 4\kappa^2 f^4$, we get
\begin{eqnarray}
 E^2&\approx&\frac{1}{2}[\epsilon_1^2+\epsilon_2^2-2\kappa f^2
\pm(\epsilon_1^2-\epsilon_2^2+\kappa f^2)] ,
\\
E^2_+&\approx& \epsilon^2_1-\frac{1}{2}\kappa f^2,
\;\;\;
E^2_- \approx  \epsilon^2_2 -\frac{3}{2}\kappa f^2 .
\end{eqnarray}
\end{description}

This simple model indicates that collectivity (collective shift
$\Delta E=E_+-\epsilon_1$) of the first RPA state is determined to a large extent by
the relative strength of the first pole.  This conclusion is confirmed
by our numerical results, see discussion of Table \ref{tab:poles}.  Thus we have
found a simple way for the prediction of the collectivity (weak or
large) of the first RPA state. In practice, it is enough to compare
the matrix elements of the first and next poles. Or, which is even easier,
one should check if the first pole fulfills the $\Delta n_z$=0
Nilsson selection rule.

\section{Comparison with QPM}

Since SRPA deals with a separable residual
interaction, this method can be directly compared with the schematic
separable QRPA exploited in QPM \cite{Sol76}. The QPM is not
self-consistent: it uses the Woods-Saxon s-p basis and its isoscalar
$\kappa_{00}$ and isovector $\kappa_{11}$ strength constants of the
residual interaction are adjusted to reproduce the experimental
energies of lowest vibrational states and giant resonances. However,
just because of the successful combination of the microscopic and
phenomenological aspects, the QPM is known to be quite accurate in
description of low-energy states. Thus it is instructive to compare
the characteristics of self-consistent models, like Skyrme QRPA, with
the relevant QPM parameters.

In this connection, let's briefly discuss the QPM strength constants
of the residual interaction and compare them with the SRPA ones. The
strength constants in the proton-neutron domain (nn, pp, np) can be
related to their counterparts in the isoscalar-isovector domain
(00,11, 01) as
\begin{eqnarray}
  \kappa_{00}&=\frac{1}{2}(&\kappa_{pp}+\kappa_{pn}+\kappa_{np}+\kappa_{nn}),
  \\
  \kappa_{11}&=\frac{1}{2}(&\kappa_{pp}-\kappa_{pn}-\kappa_{np}+\kappa_{nn}),
  \\
  \kappa_{01}&=\frac{1}{2}(&\kappa_{pp}-\kappa_{pn}+\kappa_{np}-\kappa_{nn})=\kappa_{10} .
\end{eqnarray}
The constants $\kappa_{01}=\kappa_{10}$ represent the mixing between
isoscalar (00) and isovector (11) excitations. This mixing can be
motivated by both physical (Coulomb interaction, etc) and technical
(different sizes of neutron and proton s-p basis, etc) reasons. Since
nuclei roughly keep the isospin symmetry, then
\begin{equation}
 |\kappa_{00}|, |\kappa_{11}| \gg |\kappa_{01}=\kappa_{10}| .
\end{equation}
 If to assume  $\kappa_{01}=\kappa_{10}=0$ and
$\kappa_{np}=\kappa_{pn}$, then we get
\begin{equation}\label{knn=kpp}
 \kappa_{pp}=\kappa_{nn}
\end{equation}
and the familiar QPM relations \cite{Sol76}
\begin{equation}\label{eq:kappa_00_11}
 \kappa_{00}=\kappa_{pp}+\kappa_{pn}, \qquad \kappa_{11}=\kappa_{pp}-\kappa_{pn} .
\end{equation}
>From (\ref{eq:kappa_00_11}) one gets
\begin{eqnarray}\label{eq:kappa_pp_nn}
 \kappa_{pp}=\kappa_{nn}&=&\frac{1}{2}(\kappa_{00}+\kappa_{11}),
\\
 \kappa_{pn}=\kappa_{np}&=&\frac{1}{2}(\kappa_{00}-\kappa_{11})
\end{eqnarray}
where $\kappa_{11}=\alpha \kappa_{00}$ with $\kappa_{00} > 0$.
Usually $\alpha$=-1.5 is used \cite{Sol_Shir}, which results in a
dominance of the np-interaction, $\kappa_{pn}/\kappa_{pp}$=-2.5 with
$\kappa_{pn}=\kappa_{np} > 0$ and $\kappa_{pp}=\kappa_{nn} < 0$.

For the comparison, the self-consistent SRPA calculations give
somewhat different picture. As a relevant example, the strength
constants $\kappa_{q1,q'1}=\kappa_{qq'}$ for the dominant first input
operator $r^2Y_{22}$ in $^{162}$Dy are considered.  Note that in SRPA
the relation $\kappa_{pn}=\kappa_{np}$ is kept.  SV-bas10 gives
strength constants $\kappa_{pp}, \kappa_{nn}, \kappa_{pn}>0$ with
the relations $\kappa_{pp}/\kappa_{nn}$=2.7,
$\kappa_{pn}/\kappa_{pp}$=7.7 and $\kappa_{pn}/\kappa_{nn}$=2.9.
Similar results are obtained in other nuclei.  SkM$^*$ gives
$\kappa_{nn},\kappa_{pn}>0$, $\kappa_{pp}<0$ and relations
$\kappa_{pp}/\kappa_{nn}$=-2.0, $\kappa_{pn}/\kappa_{pp}=-4.4$, and
$\kappa_{pn}/\kappa_{nn}$= 2.2.  In agreement with QPM, both forces
provide a dominant np-interaction with the proper sign.  However, in
contrast to (\ref{knn=kpp}), the weak SRPA constants $\kappa_{pp}$ and
$\kappa_{nn}$ noticeably deviate from each other, which might be a
signature of an large mixing of the isoscalar and isovector
interaction. Perhaps just this mixing, if not be properly balanced
with other parts of the interaction, partly leads to the troubles of
Skyrme QRPA with the description of $2^+_{\gamma}$ states.
A difference in sign of SV-bas and SkM$^*$ constants $\kappa_{pp}$ should be also
mentioned as demonstration of the noticeable dependence of the
residual interaction on the Skyrme force.

\end{document}